\documentstyle[oldlfont,supertabular]{l-aa}
\input psfig

\newcommand{\nat}[2]{Nat #1, #2}
\newcommand{\apj}[2]{ApJ #1, #2}
\newcommand{\apjs}[2]{ApJS #1, #2}

\newcommand{\aj}[2]{AJ #1, #2}
\newcommand{\iau}[1]{\it IUA Circ \rm No. #1}
\newcommand{\aaa}[2]{A\&A #1, #2}

\newcommand{\aeta}[2]{A\&A #1, #2}

\newcommand{\aetas}[2]{A\&AS #1, #2}

\newcommand{\mn}[2]{MNRAS #1, #2}

\newcommand{\nh}{N$_{\rm H}$}
\newcommand{\ion}[2]{\hbox{#1\,{\small #2\normalsize}}}

\newcommand{\Halp}{H${\alpha}$}

\newcommand{\Hbet}{H${\beta}$}

\newcommand{\ergcm}{erg cm$^{-2}$ s$^{-1}$}
\newcommand{\ergs}{erg s$^{-1}$}

\newcommand{\degree}{\degr}

\newcommand{\fxfo}{F$_{\rm X}$/F$_{\rm opt}$}
\newcommand{\rqvd}{r$_{90}$}

\newcommand{\cts}{cts\ s$^{-1}$}

\newcommand{\tbsp}{\rule{0pt}{10pt}}
\newcommand{\ns}{93}
\newcommand{\nsb}{87}
\newcommand{\nunid}{17}
\newcommand{\nid}{76}
\begin{document}

   \thesaurus{06         
              (13.25.3;  
               13.25.5;  
               08.01.2;  
               08.14.1;  
               08.19.1;}  
   \title{Identification of selected sources from the ROSAT Galactic Plane Survey - I
\thanks{Partly based on observations obtained at the European Southern Observatory, 
La Silla (Chile) with the 2.2m telescope of the Max-Planck-Society, with
the ESO 1.5m telescope and at the Observatoire de Haute-Provence,
CNRS, France} 
    }
 
   \author{C. Motch \inst{1}
       \and    P. Guillout \inst{2,1}
       \and    F. Haberl   \inst{2}
       \and    J. Krautter \inst{3}
       \and    M.W. Pakull   \inst{1}
       \and    W. Pietsch  \inst{2}
       \and    K. Reinsch  \inst{4}
       \and    W. Voges \inst{2}        
       \and    F.-J. Zickgraf \inst{1}
          }

   \institute{
              Observatoire Astronomique, UA 1280 CNRS, 11 rue de l'Universit\'e, 
              F-67000 Strasbourg, France
              \and 
              Max-Planck-Institut f\"ur extraterrestrische Physik, D-85740,
              Garching bei M\"unchen, Germany 
              \and 
              Landessternwarte, D-69117 Heidelberg-K\"onigstuhl, Germany
              \and 
              Universit\"ats-Sternwarte, Geismarlandstrasse 11, D-37083, G\"ottingen, Germany
              }
 
   \date{Submitted to Astronomy \& Astrophysics, Supplement Series. Received  /
Accepted }

   \offprints{C. Motch}

   \maketitle

   \begin{abstract}

We report on optical searches in the error circles of \ns \ ROSAT survey sources
located at low galactic latitudes ($|b|$ $<$ 20\degree ). These sources were extracted
from the ROSAT Galactic Plane Survey using various selection criteria on hardness
ratio, X-ray and optical brightness and integrated galactic absorption in the
direction of the source.  We find optical identifications in \nid \ cases, among which
are 25 new AGN, 6 new CVs and a new Be/X-ray binary. In order to illustrate the
relevance of the source selections applied here, we cross-correlated the ROSAT all-sky
survey bright source list with SIMBAD. Different classes of X-ray emitters populate
distinct regions of a multi dimensional parameter space involving flux ratios,
galactic latitude and \nh.  This relatively good segregation offers the possibility to
build source samples with enhanced probability of identification with a given class. 
Complete optical identification of such subsamples could eventually be used to compute
meaningful probabilities of identification for all sources using as basis a restricted
set of multi-wavelength information.  


   \end{abstract}

%

\section{Introduction}

The ROSAT all-sky survey (RASS; Voges 1992) was performed from 1990 July till 1991
February and was carried out with the X-ray telescope (XRT) and the Position Sensitive
Proportional Counter (PSPC, Pfeffermann et al. 1986). The survey mapped as many as
60,000 new sources in the soft X-ray band (0.1-2.4 keV, ML $\geq$ 10) down to limiting
fluxes of the order of a few 10$^{-13}$\,\ergcm.  Recently, the 18,811 RASS sources
having a PSPC count rate larger than 0.05\,\cts \ and ML $\geq$ 15 have been published
(Voges et al. 1996).  The study of the RASS point source content at low galactic
latitudes ($b\ \leq\ $20\degree ) is the scope of a dedicated project, the ROSAT
Galactic Plane Survey (RGPS: Motch et al.  1991).  About 15,000 RASS sources are
located in this 40\degree \ wide strip of the sky.

In the galactic plane area, already 20\% to 30\% of RGPS sources may be identified
with high confidence on the basis of positional coincidence with objects, mostly
stars, catalogued in the SIMBAD database.  Adding cross-correlation with relatively
bright stellar catalogues such as Guide Star Catalogue entries allows the
identification of about 50\% of RGPS sources (Motch et al. 1997a). 

The number of remaining unidentified sources is however still too large to allow their
systematic identification at the telescope. In order to tackle this problem, two paths
of investigations were chosen; i) selection of sources over the whole galactic sky
using criteria on their X-ray characteristics and optical content of the error box and
ii) selection of sample areas at judicious positions for complete 
optical identification.

The X-ray selected approach has led to the discovery of a number of interesting
objects, mostly soft sources which have escaped previous surveys carried out at 
higher energies and furthermore without imaging instrumentation or large sky coverage. 
Among the most noticeable results are the discovery of galactic supersoft sources
(\"Ogelman et al., 1993, Motch et al. 1994, Beuermann et al. 1995), a new class of
soft intermediate polars (Haberl \& Motch 1995), and more recently the identification
of a couple of isolated neutron stars presumably accreting from the interstellar
medium (Walter et al. 1996, Haberl et al. 1997).

On the other hand, systematic optical identification of RGPS sources in sample areas
has led to the conclusion that active coronae dominate by number the point source
content (up to 85\% of the total; Motch et al. 1997a). In contrast, the stellar
fraction identified at high galactic latitude is much lower (12-63\%; Zickgraf et al.,
1997). The modelling of this stellar population and ages derived from Lithium lines
clearly show that the stars detected in the RASS are quite young, mostly younger than
1Gyr (Guillout et al. 1996a,b). Another important result from this study is derived
from the small number of sources remaining unidentified. In particular, this implies
that the space density of isolated neutron stars accreting from the interstellar
medium is at least a factor 10 smaller than expected.

This paper is divided in two parts. First we show how X-ray and optical
characteristics of ROSAT sources may be used to build samples with enhanced
probabilities of identification with a given class of object. For that purpose we
cross-identified the ROSAT all sky survey bright source list with SIMBAD. Second, we
report on optical observations of \ns \ selected RGPS sources. The source list
contains various kinds of X-ray selected sub samples out of which several
identifications not repeated here were already published. 

Our identification programme of X-ray selected sources in the galactic plane is
continuing and we expect to provide more RGPS source identifications in a future paper.

\section{Cross-identification of the ROSAT all-sky survey bright source catalogue with
SIMBAD}

\subsection{BRASS-SIMBAD correlation}

Studying the clustering properties of identified X-ray sources in a multi-wavelength
parameter space can help finding domains in which the probability of association with
a given class of object is higher than on the average.  For instance, at low galactic
latitude the overwhelming domination of stellar coronae makes the discovery of new CV
a difficult task as they only account for $\approx$ 1\% of the total number of sources
(Motch et al.  1996) whereas their frequency is much higher among either hard, or soft
and bright selected sources.

If completely identified flux limited samples are obtained using catalogue
cross-correlation or work at the telescope, then the probability that X-ray emission
from a given source originates from a particular class of emitters (star, AGN, etc.)
may be eventually estimated in a quantitative manner. This allows the computation of a
statistical identification using as input the restricted set of multi-wavelength
parameters available for all sources.  

RGPS sources identified in SIMBAD are far from making a complete sample. SIMBAD is
roughly optical flux limited, although at very different levels depending on the
nature, stellar or extragalactic, of the object. As most of the recent additions are
coming from the literature, the whole collection is highly heterogeneous. Clearly,
the set of SIMBAD identifications cannot be used to provide real probabilities of
identification.  However, it may be used to construct selected samples of presumably
enhanced probabilities.  

A first analysis was made in 1991, using X-ray and optical characteristics of a limited
number of stars, AGN, CVs and accreting binaries identified in preliminary and partial
releases of the RASS. The selection criteria used to build the source samples studied
in this paper are derived from this early analysis.

The recently published ROSAT all-sky survey bright source catalogue (BRASS) by Voges
et al. (1996) constitutes a sample of larger size and of much better quality than the
one originally used in 1991. Therefore, for the sole aim of illustrating in an
extensive manner the X-ray selection criteria used in this work, we decided to
cross-correlate BRASS with SIMBAD. We underline that the associations between bright
RASS sources and SIMBAD entries used in this paper have not been human screened and
cannot be considered individually as real identifications. Nevertheless their overall
statistical properties may still be used for our purpose.  

The BRASS covers the whole sky whereas the RGPS is by definition limited to the
galactic plane area. However, selection rules arising from the BRASS catalogue can
usefully be applied to the restricted RGPS area. The variability of parameters with
galactic latitude or \nh \ can also be better studied using the whole sky catalogue.

The first step of the cross-correlation consisted in extracting from SIMBAD all
objects located within 3\arcmin \ from the ROSAT source, retrieving the optical position
and associated error, the first identifier, possible ROSAT name, optical magnitudes,
spectral types, full object type, number of references and Einstein measurements when
available. In a second step, we retained from the SIMBAD cross correlation log entries
having a non null intersection between the ROSAT 90\% confidence circle and the SIMBAD
error circle. As the error radius of Einstein sources is underestimated in SIMBAD, we
updated it manually to 50\arcsec . When several SIMBAD objects remained, we retained
the one closest to the X-ray position.  With a mean 90\% error radius of 27\arcsec \
for BRASS sources, and a total number of $\sim$1.46 10$^{6}$ SIMBAD objects having in
most cases small positional errors ($\leq$ 1\arcsec),  we expect about 120 spurious
matches between a SIMBAD and a ROSAT source. This estimate ignores effects resulting
from inhomogeneities in the surface density of SIMBAD and BRASS entries.  We list in
Table \ref{simcros} some statistics of the cross-correlation.  

\begin{table}
\caption{BRASS-SIMBAD correlation}
\label{simcros}
\begin{tabular}{lrr}
Sample      & Number  & fraction of        \\
            &         & Total BRASS (\%) \\
\hline
Unidentified & 10705  & 56.9 \\
1 SIMBAD match & 6123 & 32.6 \\  
2 SIMBAD matches & 1313 & 6.9 \\
$\geq$3 SIMBAD matches & 670 & 3.6 \\
\hline
\end{tabular}
\end{table}

Clearly, the large majority of SIMBAD matches are real since only $\sim$ 1.5\% of
the sources with one or more SIMBAD entries are expected to be spurious.  Considering
the restrictive aim of this work, we did not investigate in detail the reasons
explaining the relatively large number of sources with multiple SIMBAD matches.
Probably, the high surface density of SIMBAD entries in some specific small regions of
the sky, multiple stellar systems and overall the fact that a number of astrophysical
objects still appear as distinct SIMBAD objects could account for this effect.

Most of the proposed SIMBAD identifications of BRASS sources (61\%) are with stars and
only 20\% with external galaxies and AGN. The high identified stellar fraction
obviously reflects the emphasis put on stars in SIMBAD. 
 
\begin{figure}
\psfig{figure=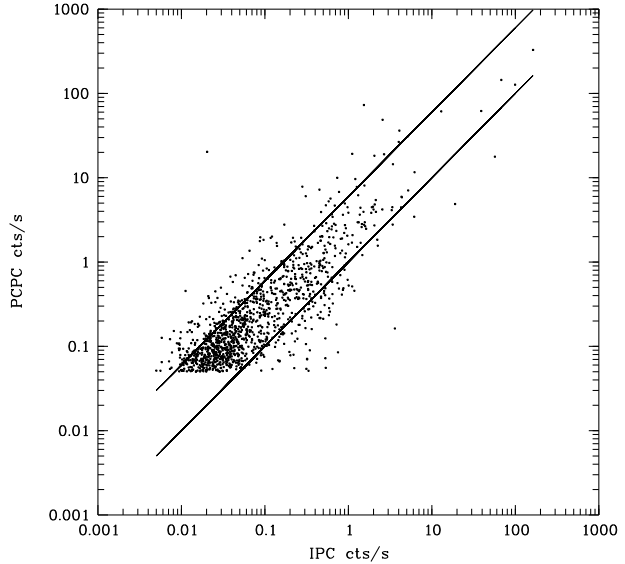,width=8.8cm,bbllx=1.0cm,bburx=22cm,bblly=1.0cm,bbury=20cm,clip=true}
\caption[]{The relation between Einstein IPC and PSPC count rates for 1244
Einstein sources identified with BRASS entries. The two lines represent the
relation expected for soft and hard sources.}
\label{ipc_pspc}
\end{figure}

The quality of the cross-correlation may be judged from Fig. \ref{ipc_pspc} which
shows the relation between Einstein IPC and ROSAT PSPC count rates for the 1244
Einstein sources recovered in the BRASS. The ratio of the count rates of the two
instruments strongly depends on source spectrum with an additional scatter due to
variability. We show for comparison two lines (PSPC = 1 $\times$ IPC and  PSPC = 6
$\times$ IPC corresponding to hard (power law photon index = 0) and soft (thermal
spectra with logT = 5.8) sources respectively. As expected, most sources lie between
these two extreme relations. The outstanding source at IPC = 0.02\,\cts \ and PSPC =
20\,\cts \ is the low-mass X-ray binary transient EXO 0748-676 which was detected by
Einstein in the low state before its discovery by EXOSAT in outburst.

\subsection{X-ray and optical properties of identified BRASS sources}

For source classification purposes, the most interesting parameters are flux ratios in
various energy bands, including the conventional X-ray hardness ratios, but also \fxfo
\ ratios as well as optical colours.  Another important information for galactic
studies is the line of sight absorption which can be estimated from \ion{H}{I} and CO
data.

Arguments involving the \fxfo \ ratio have been used by many authors in the context of
the Einstein Observatory surveys at high and low galactic latitudes (e.g., Maccacaro
et al., 1982;  Hertz \& Grindlay 1984). ROSAT offers a slightly improved spectral
response compared to Einstein and allows PSPC hardness ratios to be used as additional
information, at least for relatively bright sources.

\subsubsection{Hardness ratios}

\begin{figure*}

\begin{tabular}{cc}

\psfig{figure=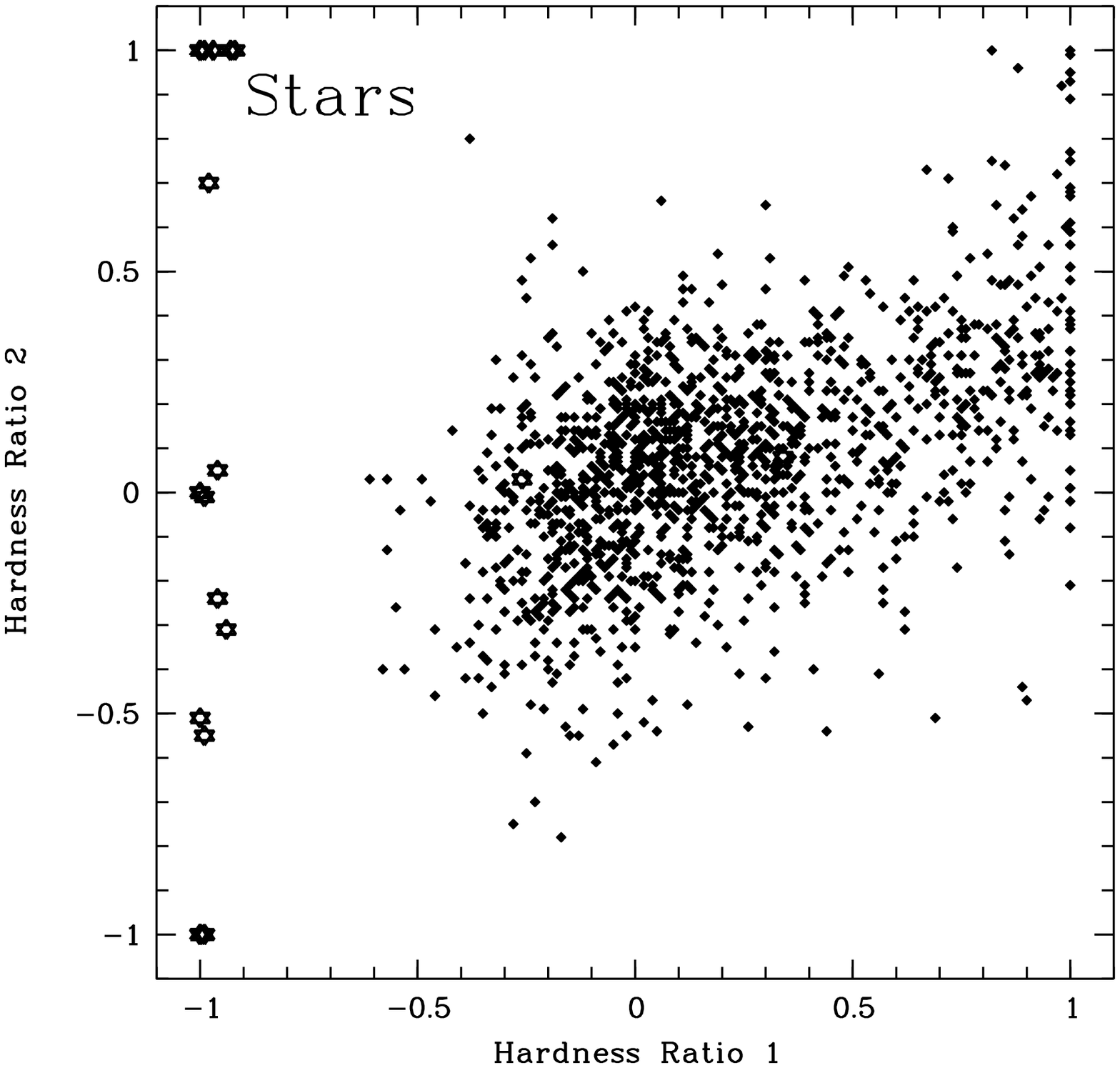,width=8.8cm,bbllx=0.0cm,bburx=22cm,bblly=0.0cm,bbury=22cm}

&
\psfig{figure=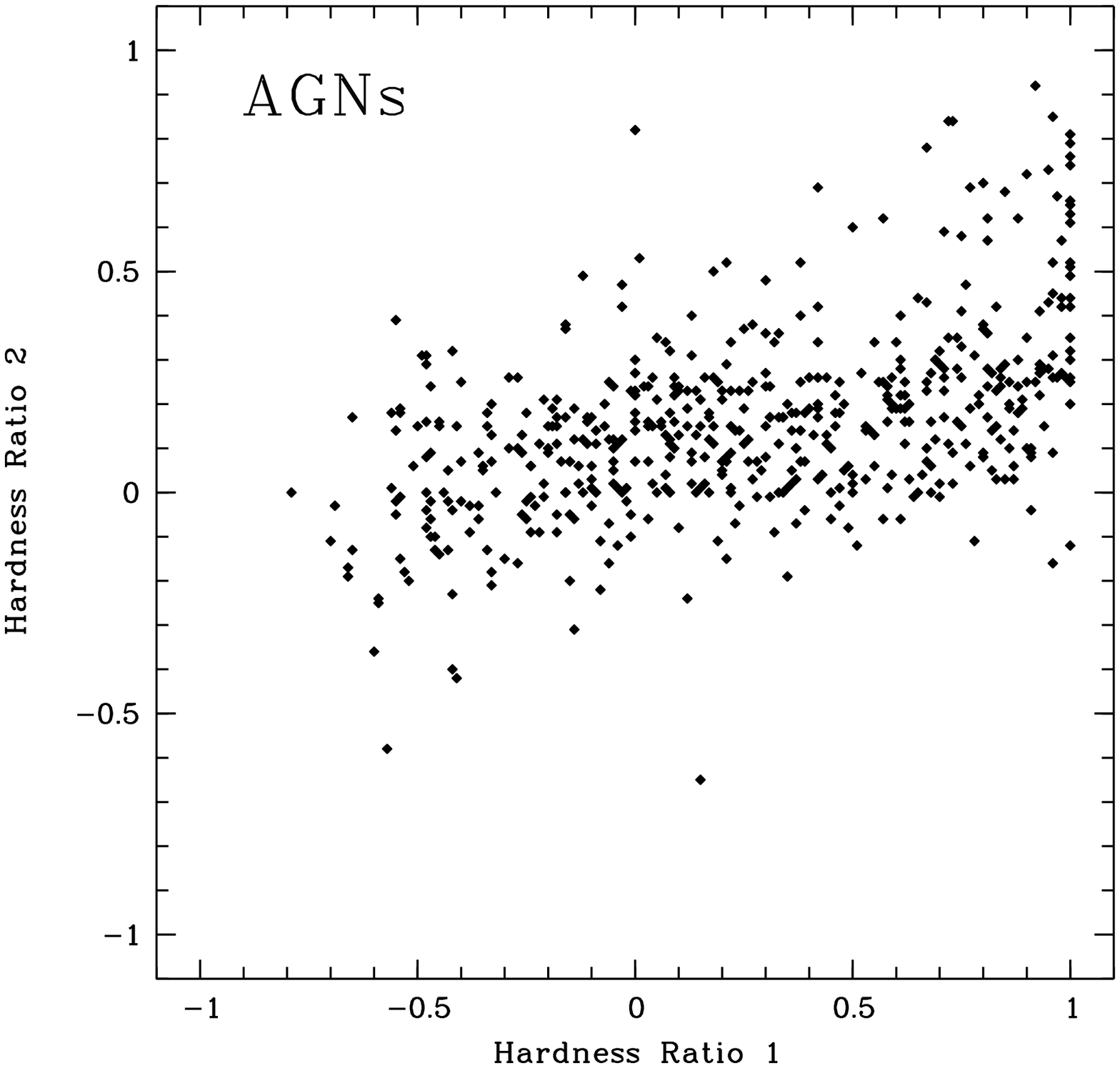,width=8.8cm,bbllx=0.0cm,bburx=22cm,bblly=0.0cm,bbury=22cm}

\\

\psfig{figure=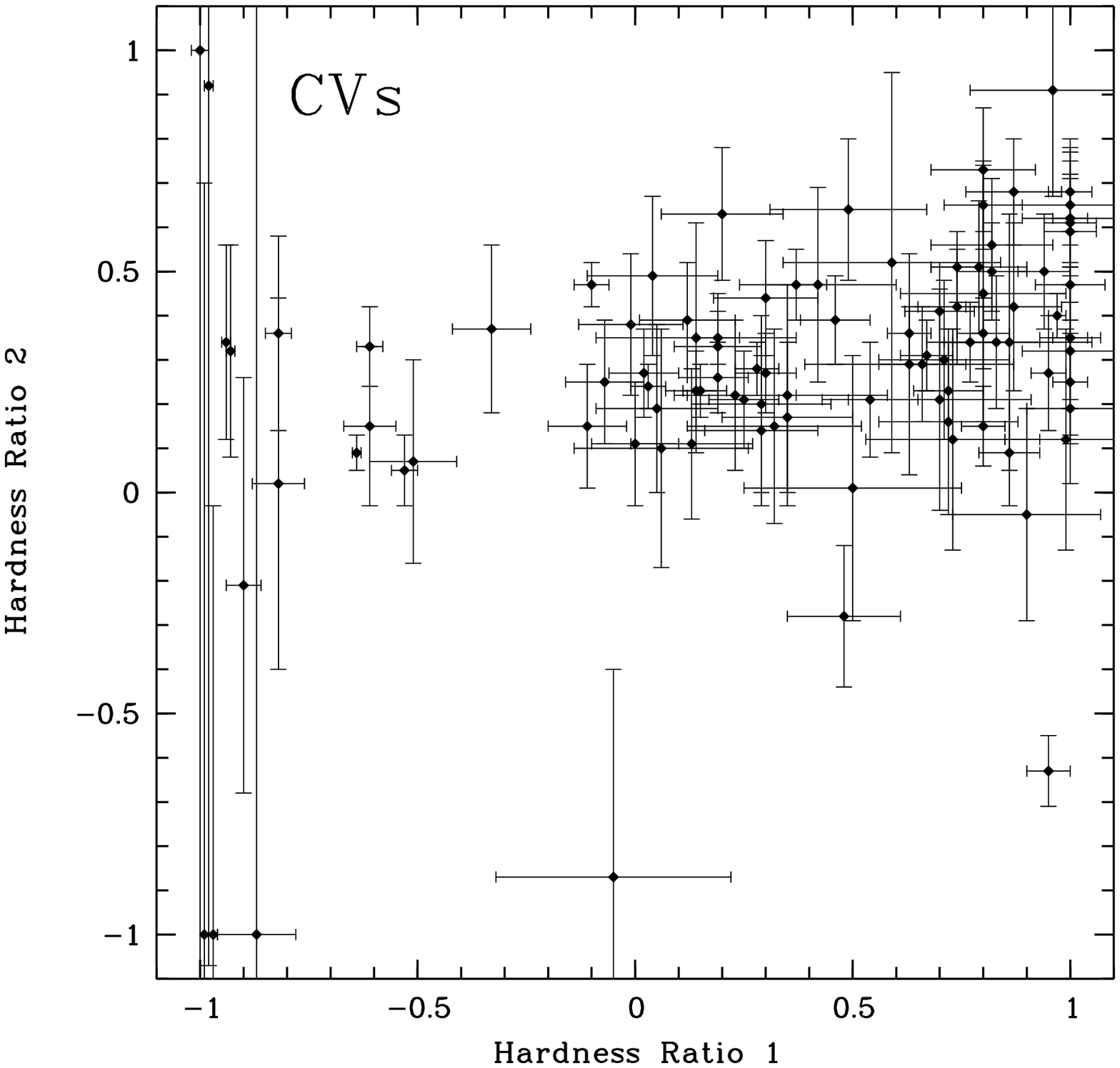,width=8.8cm,bbllx=0.0cm,bburx=22cm,bblly=0.0cm,bbury=22cm}

&

\psfig{figure=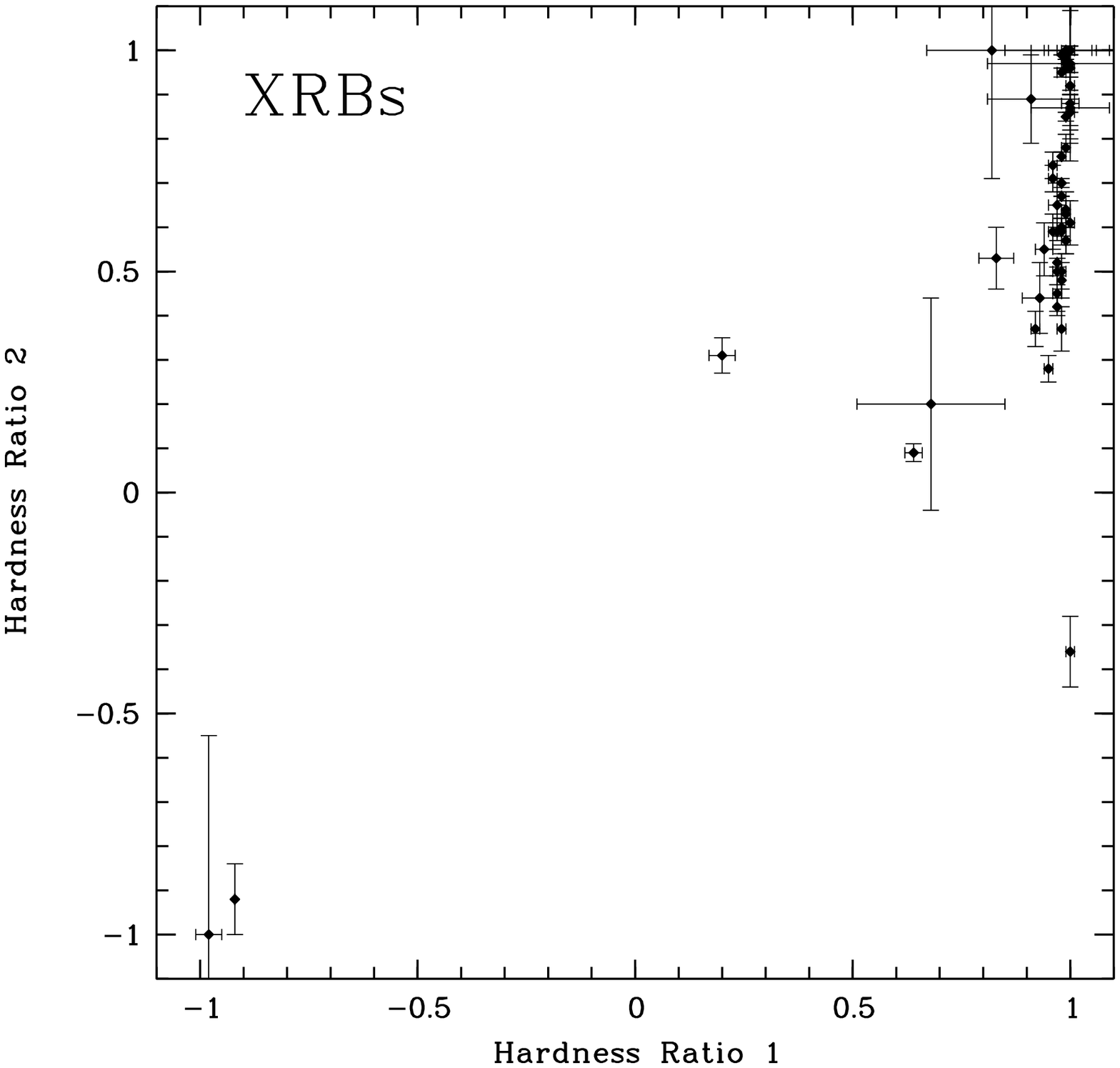,width=8.8cm,bbllx=0.0cm,bburx=22cm,bblly=0.0cm,bbury=22cm}

\\

\end{tabular}

\caption[]{Position of various classes of X-ray emitters in the HR1 HR2 diagram. 
Asterisks in the Stars diagram represent white dwarfs. The scatter in HR2 of white
dwarfs is due to the large error on this ratio for such soft sources. For
non-degenerate stars and AGN, we only show sources with errors less than 0.2 on both
ratios.} 

\label{all_h1h2}

\end{figure*}

We plot in Fig. \ref{all_h1h2} the positions in the hardness ratio diagram of BRASS
sources cross-identified in SIMBAD with different classes of objects.

The exact energy range used for the computation of hardness ratios changes with the
version of the Scientific Analysis System Software (SASS; Voges et al., 1992). 
Whereas the \ns \ sources discussed in this paper were the output of SASS-I (see
section 3.1), the BRASS catalogue (Voges et al. 1996) uses SASS-II processing in which
the hardness ratios are defined as:

\begin{displaymath}
{\rm HR1}\ = \frac{(0.5-2.0)-(0.1-0.4)}{(0.1-0.4)+(0.5-2.0)} \ {\rm (SASS-II)}
\end{displaymath}
\begin{displaymath}
{\rm HR2}\ = \frac{(1.0-2.0)-(0.5-1.0)}{(1.0-2.0)} \ {\rm (SASS-II)}
\end{displaymath}
\noindent where (A-B) is the raw background corrected source count rate in the A$-$B
energy range expressed in keV.

Active coronae populate the central part of the HR1/HR2 diagram with some tail
extending towards hard spectra. Stellar coronae are known to exhibit a range of
temperatures between 3 10$^{6}$\,K and 10$^{7}$\,K, with the most active and luminous
stars also exhibiting the highest kT and hardest spectra. In this diagram, isolated
white dwarfs are all found at HR1 $\leq$ $-$0.9. At the BRASS flux level, the vast
majority of active coronae are located within 100 pc from the Sun (e.g. Guillout et
al.  1996b) and except for the early type stars and some of the most luminous late
type binaries, the effects of interstellar absorption are negligible in both X-ray
colours.

In general, the AGN found in SIMBAD do not exhibit HR2 smaller than $\approx$~$-$0.2
but have a large scatter in HR1 which is closely related to galactic foreground
absorption (see below). However, ROSAT has discovered some AGN with very steep 
spectra which are so far barely represented in SIMBAD (e.g. Greiner et al. 1996).

The bulk of the cataclysmic variables populate the upper right quadrant, avoiding very
hard HR2 values.  Among outstanding sources we find at HR1 = 0.95$\pm$0.05 and HR2 =
$-$0.63$\pm$0.08 the peculiar soft IP candidate RX J1914.4+2456 which exhibits strong
interstellar absorption (Haberl \& Motch, 1995; Motch et al. 1996).  The supersoft
X-ray emission from GQ Mus = Nova Muscae 1983 at HR1 = $-$0.05$\pm$0.27 and HR2 =
$-$0.87$\pm$0.47 was discovered by \"Ogelman et al. (1993).  Cataclysmic variables with
HR1 $\leq$ 0.1 are all polar or soft intermediate polar systems.

Most X-ray binaries exhibit very hard X-ray hardness ratios resulting both from the
usually high interstellar absorption towards these remote sources and from an
intrinsically hard spectrum (kT $\geq$ 2 keV with sometimes local photoelectric
absorption such as in Be/X-ray binaries for instance). The two objects at the lower
left corner are supersoft sources in the Large Magellanic Cloud.  The galactic
supersoft source RX J0925.7-4758 (HR1 = 1.00$\pm$0.01 and HR2 = $-$0.36$\pm$0.08)
exhibits an intrinsically soft spectrum affected by strong interstellar absorption. 
The source at HR1 = 0.64$\pm$0.02 and HR2 =  0.09$\pm$0.02 is the LMC transient 1A
0538-66 and that at HR1 = 0.20$\pm$0.03 and HR2 = 0.31$\pm$0.04 is the peculiar M
giant binary HD 154791/A 1704+241.

\subsubsection{Hardness ratios and \fxfo }

\begin{figure*}
\begin{tabular}{cc}

\psfig{figure=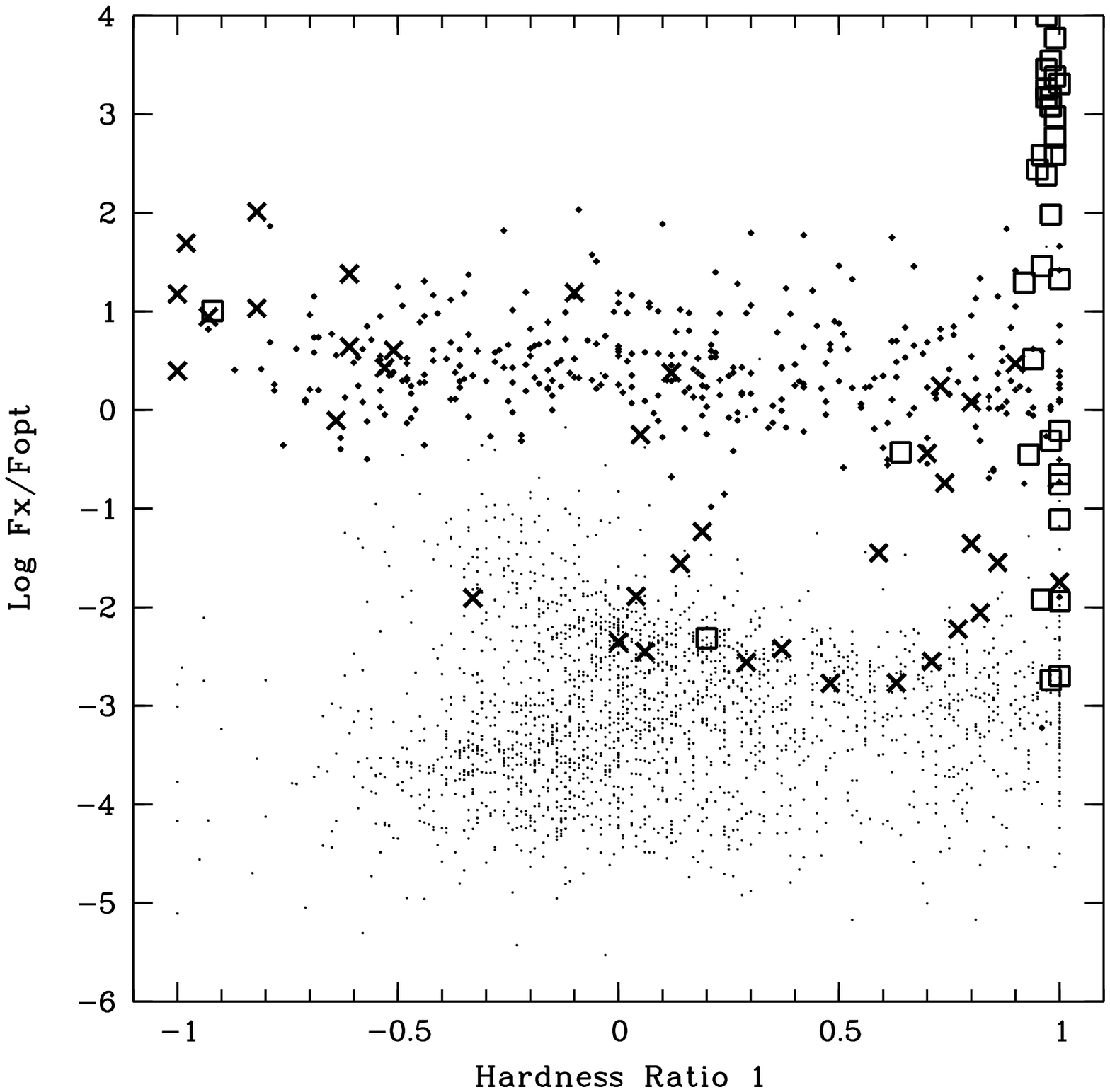,width=8.8cm,bbllx=0.0cm,bburx=22cm,bblly=0.0cm,bbury=22cm}

&

\psfig{figure=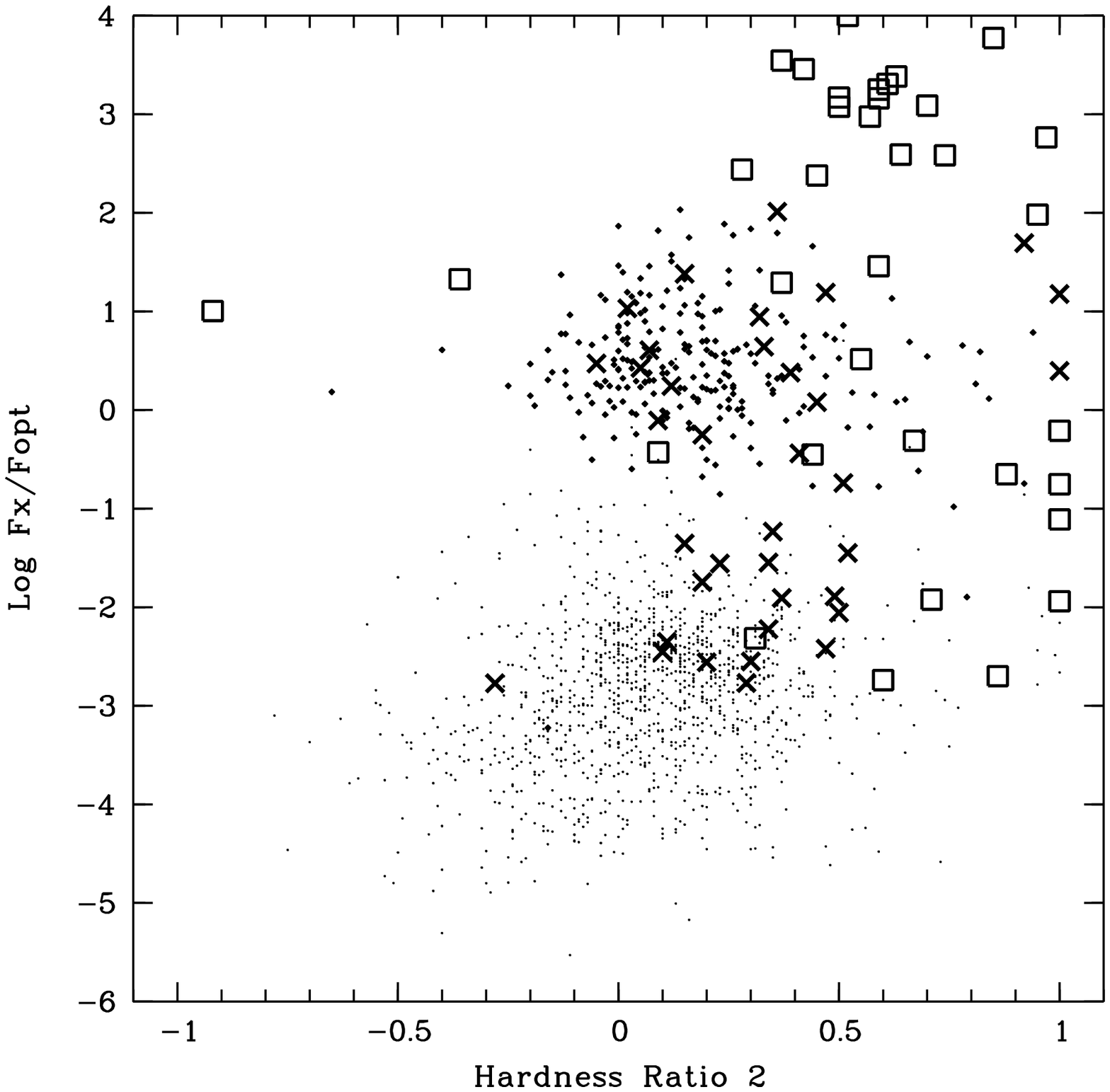,width=8.8cm,bbllx=0.0cm,bburx=22cm,bblly=0.0cm,bbury=22cm}

\\
\end{tabular}

\caption[]{The position of various classes of identified BRASS sources in the HR1
(left) or HR2 (right) / Log(\fxfo ) diagram. Small dots represent stars, large dots
AGN, squares are X-ray binaries and crosses are cataclysmic variables. For stars and
AGN only sources having errors on HR1 smaller than 0.2 are plotted}

\label{ratio_12}

\end{figure*}

The most informative diagrams are without doubt those involving optical information,
basically in the form of the \fxfo \ ratio. The X-ray to optical flux ratio can be
defined as log(\fxfo ) = log(PSPC count rate) + V/2.5 $-$ 5.63, following the
expression used by Maccacaro et al. (1982) for the Einstein medium sensitivity survey
and assuming an average energy conversion factor of 1 PSPC\,\cts \ for a 10$^{-11}$ erg
cm$^{-2}$ s$^{-1}$ flux in the range of 0.1 to 2.4 keV. We show in Fig. \ref{ratio_12}
the position of stars, AGN, X-ray binaries and cataclysmic variables in the X-ray
colour versus \fxfo \ ratio diagram.

Although stars and AGN have similar X-ray colours, their mean X-ray to optical ratios
are obviously quite different and the two populations are well separated in the HR1/2
\fxfo \ diagram. X-ray binaries are essentially recognizable from their hard X-ray
spectra and usually large \fxfo . The low \fxfo \ X-ray binary tail consists of high
mass X-ray binaries. Cataclysmic variables exhibit a large range of X-ray colours and
\fxfo \ ratios and can be somewhat confused with both the AGN and the most active part
of the stellar population. However, the addition of a B-V or U-B optical index would
allow to distinguish further between these overlapping populations.

\subsubsection{Interstellar absorption}

\begin{figure*}

\begin{tabular}{ccc}

\psfig{figure=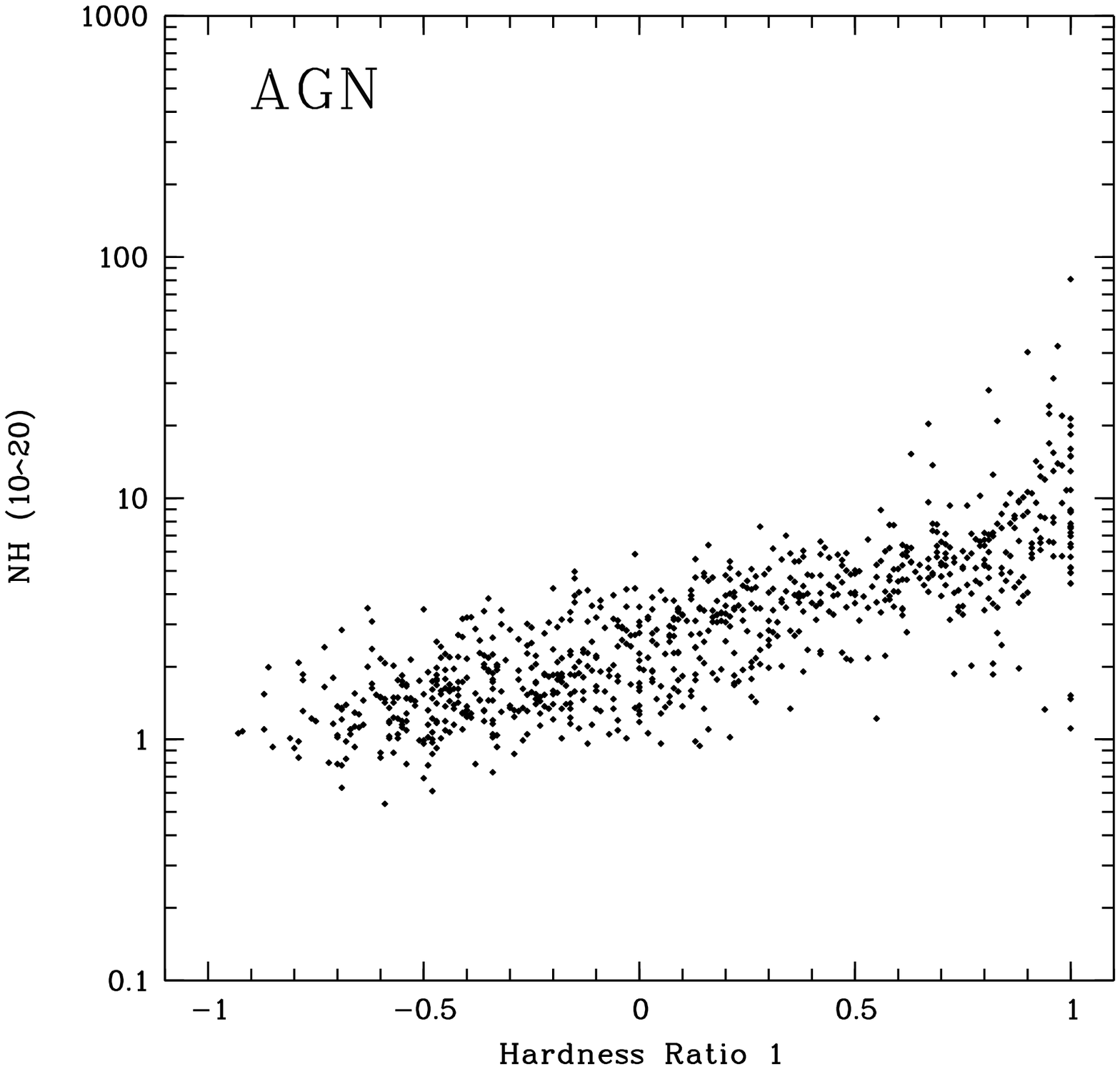,width=5.8cm,bbllx=0.0cm,bburx=22cm,bblly=0.0cm,bbury=22cm}

&

\psfig{figure=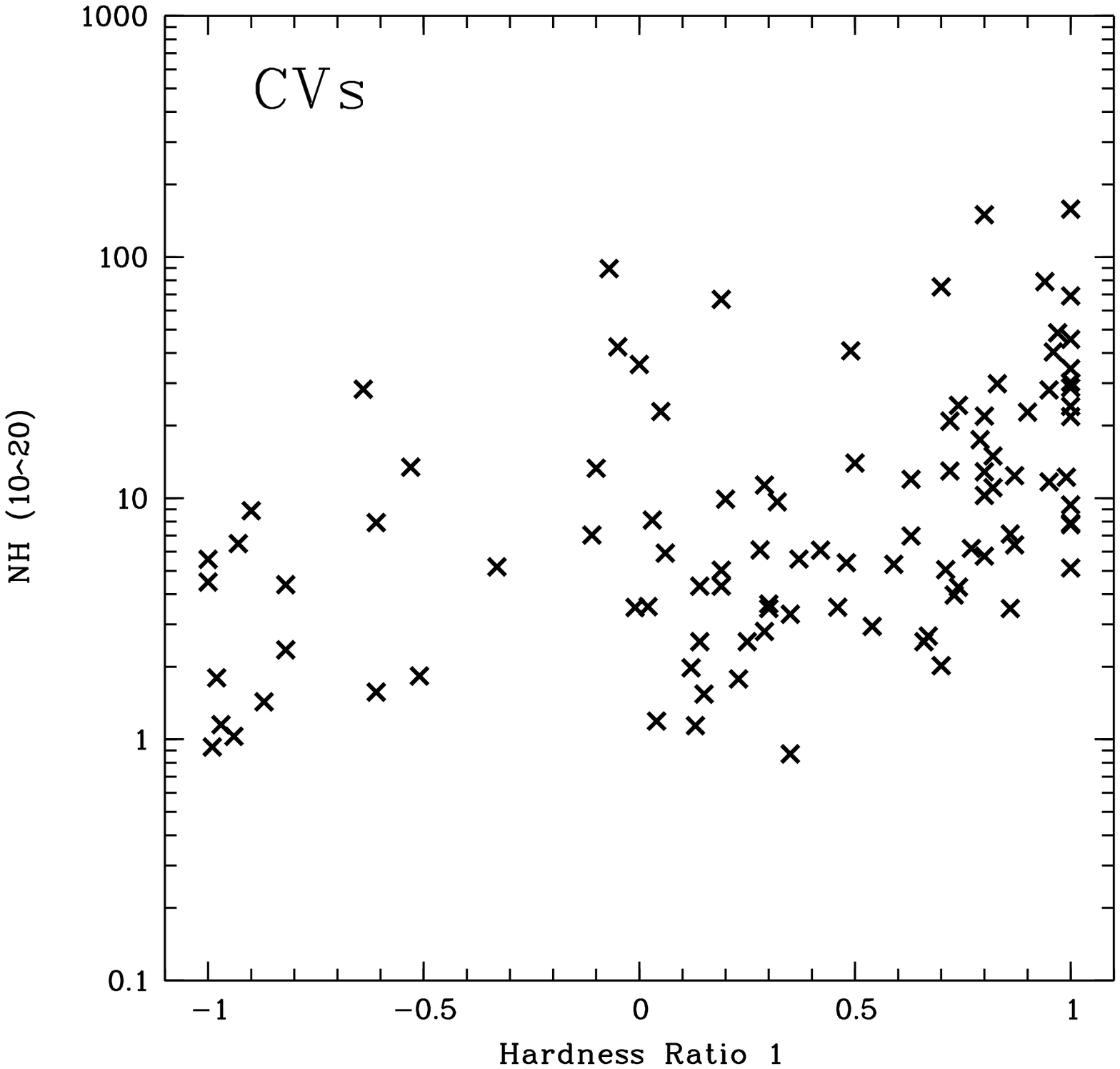,width=5.8cm,bbllx=0.0cm,bburx=22cm,bblly=0.0cm,bbury=22cm}

&

\psfig{figure=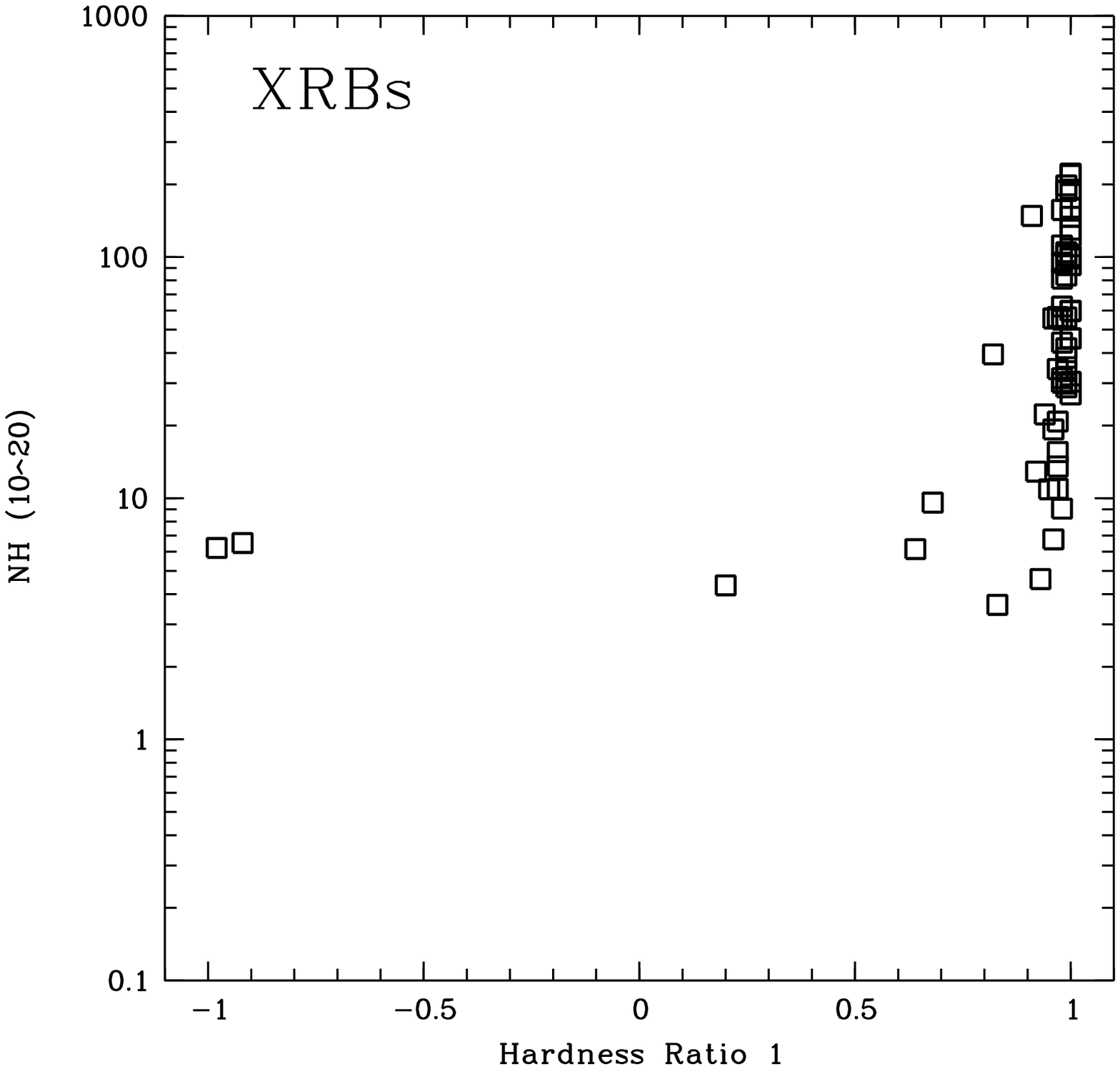,width=5.8cm,bbllx=0.0cm,bburx=22cm,bblly=0.0cm,bbury=22cm}

\\

\end{tabular}

\caption[]{Variation of HR1 with the integrated galactic HI column density for
various classes of X-ray emitters. For stellar and AGN identifications, only
sources with error on HR1 smaller than 0.2 are plotted}

\label{hr1_nh}

\end{figure*}

\begin{figure*}

\begin{tabular}{ccc}

\psfig{figure=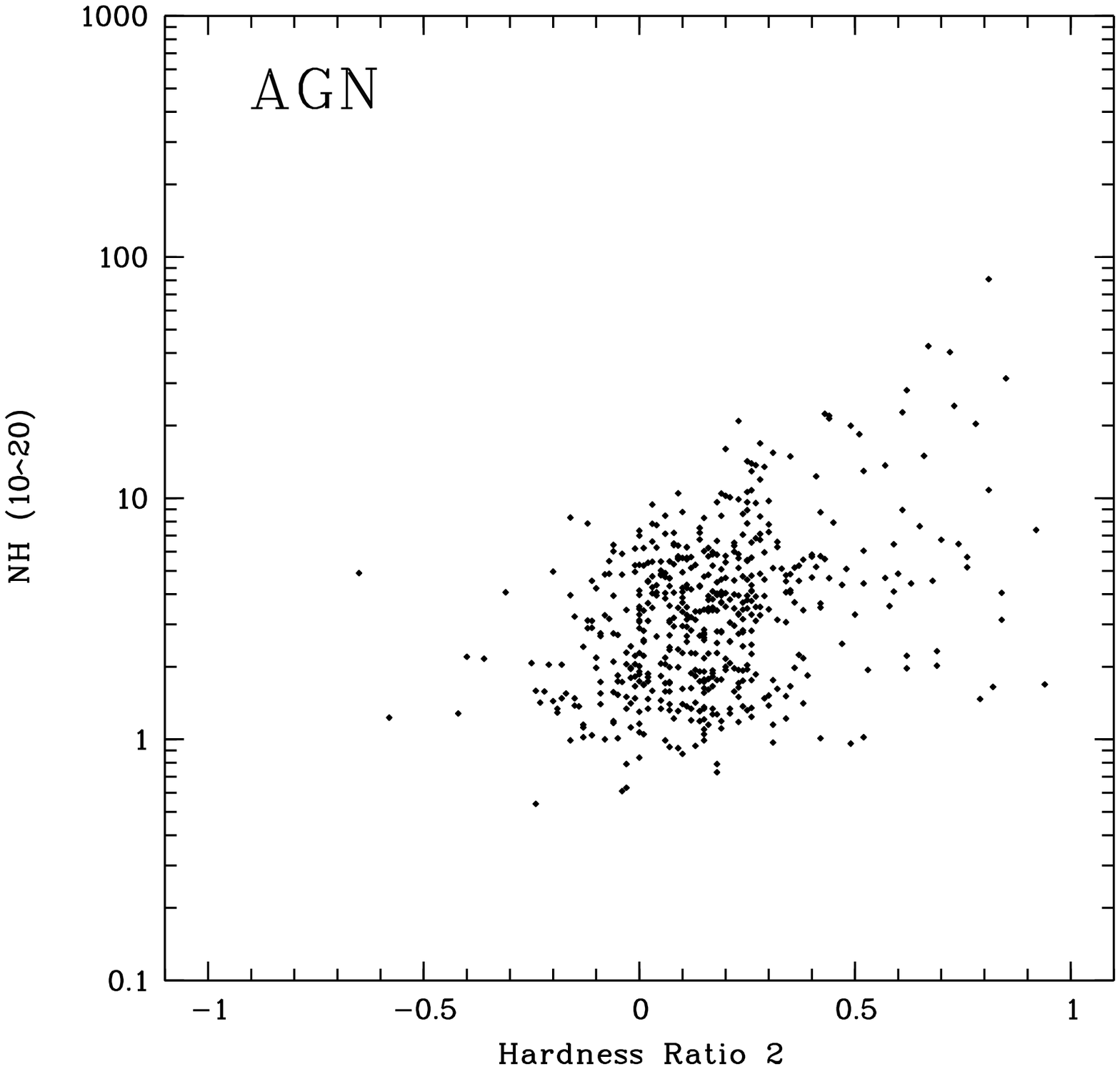,width=5.8cm,bbllx=0.0cm,bburx=22cm,bblly=0.0cm,bbury=22cm}

&

\psfig{figure=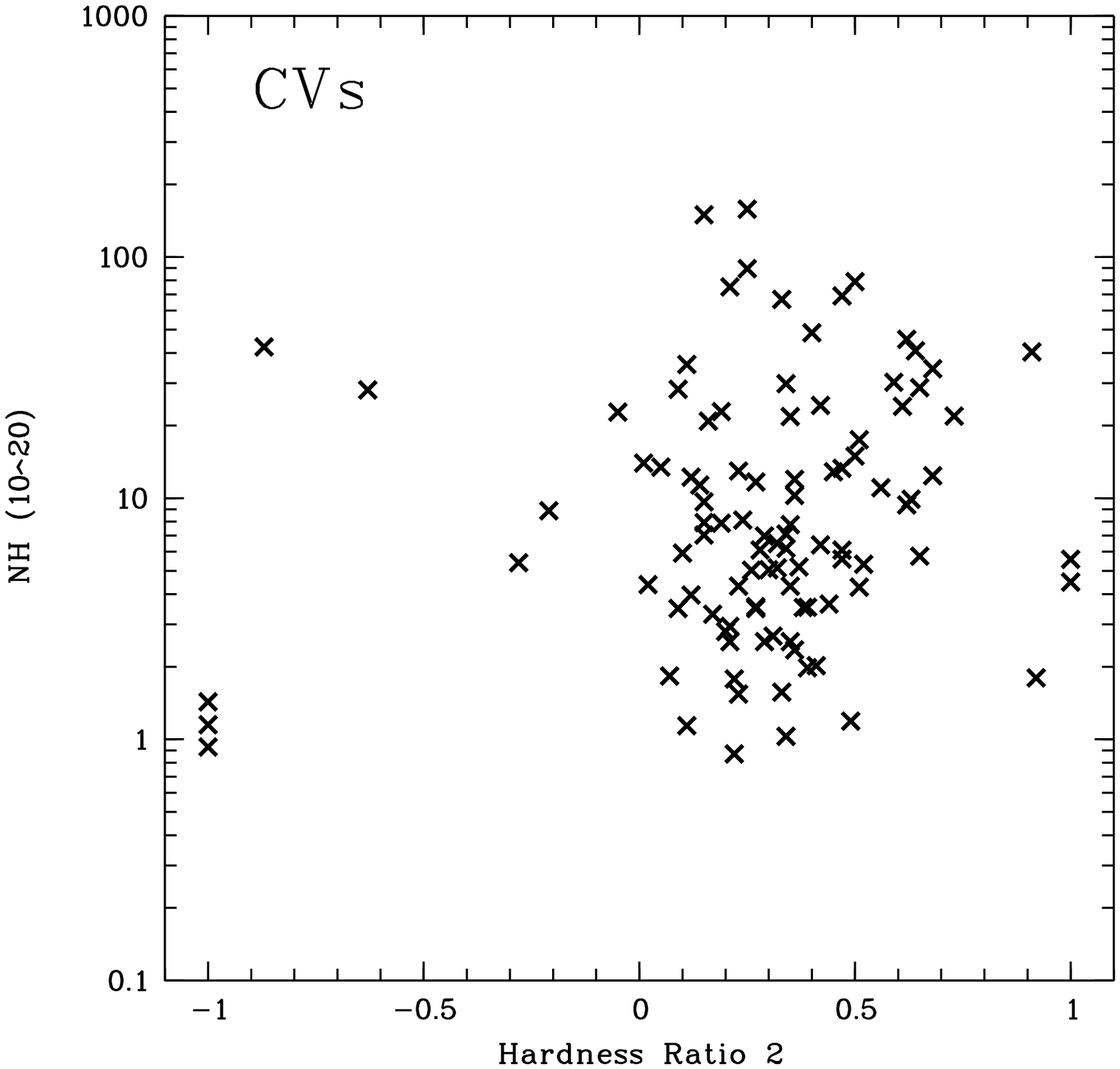,width=5.8cm,bbllx=0.0cm,bburx=22cm,bblly=0.0cm,bbury=22cm}

&

\psfig{figure=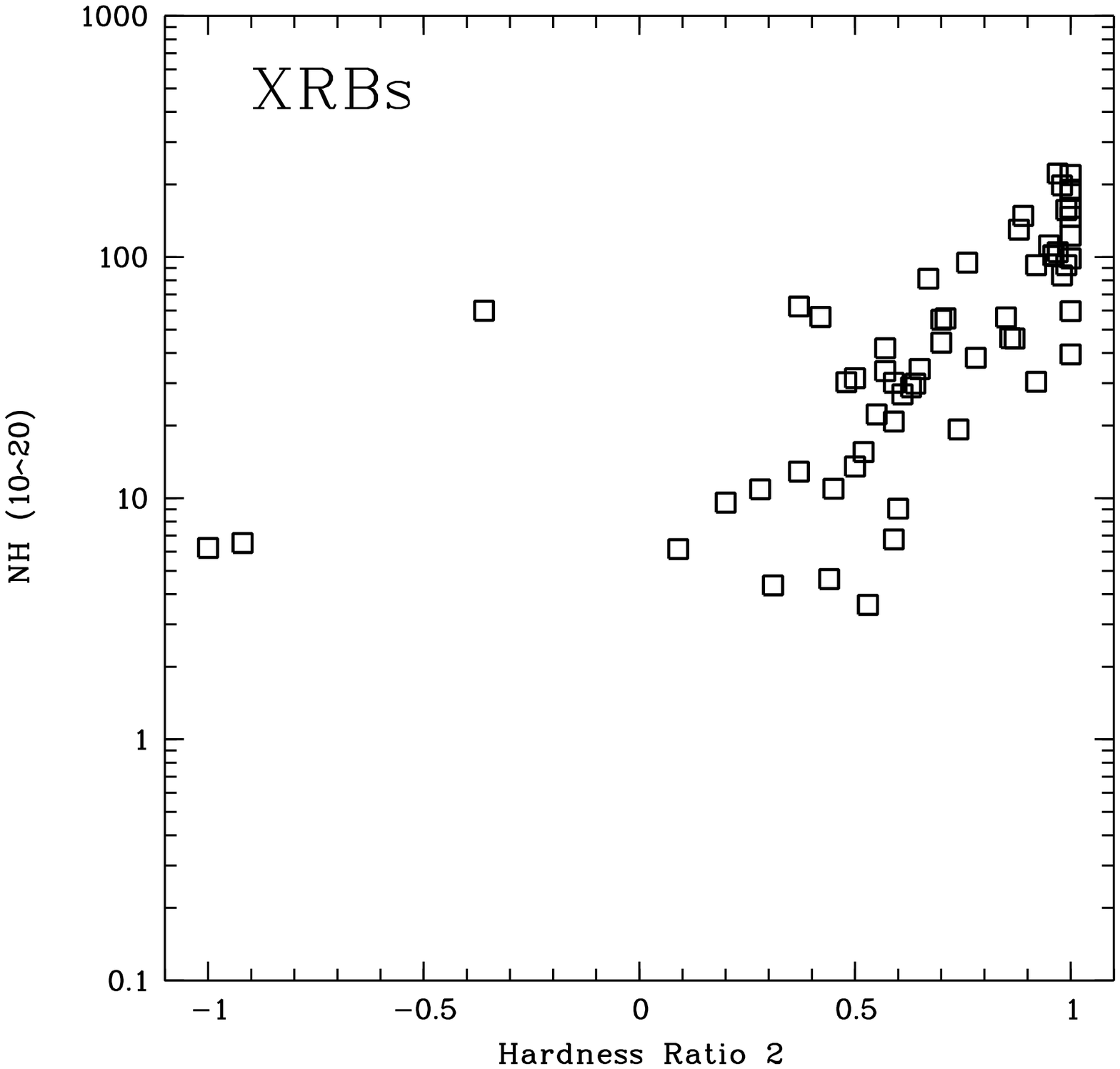,width=5.8cm,bbllx=0.0cm,bburx=22cm,bblly=0.0cm,bbury=22cm}

\\

\end{tabular}

\caption[]{Variation of HR2 with the integrated galactic HI column density for
various classes of X-ray emitters. For stellar and AGN identifications, only
sources with error on HR1 smaller than 0.2 are plotted}

\label{hr2_nh}

\end{figure*}

An efficient way to discriminate between local and remote populations of X-ray sources
is to use the anisotropy produced by the Galaxy, essentially in terms of scale height
and interstellar absorption. This effect is illustrated in Figs. \ref{hr1_nh} and
\ref{hr2_nh} which display the position of AGN, CVs and X-ray binaries in the
integrated HI / hardness ratio diagram.

Most active stars detected in the RASS are little affected by interstellar absorption
and their X-ray colours do not vary with galactic latitude (or integrated column
density). 

Apparently, the shape of the low energy spectrum of AGN is rather constant with the
consequence that hardness ratio 1 is well correlated with the integrated HI column
density as shown in Fig. \ref{hr1_nh}. Actually, this relation is well defined and we
used it for preselecting AGN candidates with a high rate of success. However, the HR2
distribution is relatively peaked around a value of 0.1  and does not vary with \nh. 
This is probably due to a selection effect against highly absorbed AGN which are
likely to be missing in the mostly optically selected SIMBAD sample.  

With X-ray luminosities up to 10$^{32-33}$ \ergs, cataclysmic binaries can
be detected to distances as large as 1 kpc or more at the BRASS sensitivity. There is
indeed a slight tendency that cataclysmic variables exhibiting the hardest HR1 are
preferentially found at low galactic latitudes but the large variety of spectra
emitted in this class (from very soft polars to intrinsically absorbed intermediate
polars) somewhat blurs the picture.

In contrast, luminous X-ray binaries are seen by ROSAT at very large distances in
deeply absorbed regions of the galactic plane and apart from a few cases (e.g. 
supersoft sources), their hardness ratio HR1 is close to +1, indicating that all X-ray
photons below 0.4 keV are blocked by interstellar absorption. Effects of interstellar
absorption are also seen on hardness ratio HR2 which involves higher energy bands. Not
surprisingly, Fig.  \ref{hr2_nh} shows that sources lying in the most absorbed
direction of the Galaxy (and therefore lower galactic latitudes) are also the hardest
in HR2.  

\section{Identification of sources from the ROSAT Galactic Plane Survey}

\subsection{The RGPS source list}

The \ns \ sources discussed in this paper were extracted from the RGPS source list
using selection criteria described in section 3.5 below.  The RGPS source list results
from the survey reduction performed by SASS-I in October 1991.  This RASS reduction
has well known documented flaws, namely a PSPC count rate overestimated by a factor of
$\approx$ 1.2, the appearance of some ghost sources caused by problems in the
satellite attitude reconstruction in spinning mode and uneven sensitivity due to the
strip accumulation.

Spacecraft problems prevented the detection of sources between ecliptic longitudes
41\degree \ and 49\degree \ and ecliptic longitudes 221\degree \ and 229\degree \
leaving about 5\% of the galactic sky without coverage.  

The 1991 SASS-I version stacked photons in 2\degree \ wide strips along the
great scan circles resulting in frequent multiple detections of the same source in
adjacent strips. The lists of sources derived from each strip were merged into a
single master list totaling about 15,000 sources at $|b|$ $\leq$ 20\degree . This list
constitutes the database for the RGPS. Details on the merging procedure can be found
in Motch et al. (1997b).

Errors on ROSAT X-ray positions are quadratic sums of the statistical uncertainty with
which the centroid of the X-ray image is positioned on the pixel grid by the Maximum
Likelihood source detection algorithm and of a systematic attitude error, estimated to
be of the order of 8\arcsec \ (Motch et al. 1997b).

Among the \ns \ sources, a total of \nsb \ are bright enough to be also found in the
RASS bright source catalogue (1 RXS, Voges et al. 1996) discussed in the previous
section. The SASS process used for the production of the bright source list (SASS-II)
detects sources on square sky areas, eliminating thus the uneven sensitivity resulting
from the strip approach used in 1991. Updated attitude reconstruction removed ghost
sources sometimes present in the 1991 version and human data screening ensured high
quality.  Most RGPS and BRASS positions are fully consistent with a median difference
in position of 7.6\arcsec .  The 90\% confidence radii are similar.  In a few cases
mentioned below in the notes on individual objects, the offset between the RGPS and
BRASS positions is significantly larger than \rqvd \ probably because of the use of an
improved attitude solution in the SASS-II reduction.  The 1991 SASS-I reduction also
uses slightly different energy ranges than the SASS-II reduction for computing
hardness ratios 1 and 2:

\begin{displaymath}
{\rm HR1}\ = \frac{(0.40-2.40)-(0.07-0.40)}{(0.07-2.40)} \ {\rm (SASS-I)} 
\end{displaymath}
\begin{displaymath}
{\rm HR2}\ = \frac{(1.00-2.40)-(0.40-1.00)}{(0.40-2.40)} \ {\rm (SASS-I)}
\end{displaymath}

\noindent where (A-B) is the raw background corrected source count rate in the A$-$B
energy range expressed in keV. There is no one to one relation between SASS-I and
SASS-II hardness ratios as their values depend on the details of the observed count
distribution in energy. However, a meaningful mean relation exists which can be used
to propagate the ranges from one reduction to the other.

\subsection{Optical observations}

Optical material was collected at the Observatoire de Haute-Provence, CNRS, France for
northern sources and at ESO, La Silla for the southern sky. The present observational
material was acquired during several runs dedicated to the identification of ROSAT
galactic plane survey sources in general which were carried out from 1991 till 1995 by
various observers.  

At OHP, multicolour CCD imagery was usually obtained with the 1.2\,m telescope few days
before the spectroscopic run with the 1.9\,m. In most cases we used  B and I band
filters with additional U band exposures in a few instances. Pixel size was 0.85 or
0.77 arcsec on the sky depending on the CCD chip used. At the 1.9\,m telescope, we
operated the CARELEC spectrograph (Lemaitre et al.  1990). Most of the time we used
two dispersions, a low resolution mode 260 \AA /mm ($\lambda \lambda$ 3500 $-$ 7500
\AA ; FWHM resolution $\approx$ 14 \AA ) and a medium resolution mode in the blue 33
\AA /mm ($\lambda \lambda$ 3800 $-$ 4300 \AA ; FWHM resolution $\approx$ 1.8 \AA ).  

ESO data were acquired at the occasion of four runs in May 1991, April 1992, February
1994 and February 1997. In 1991, 1994 and 1997, we used the Boller \& Chivens
spectrograph at the ESO 1.5m telescope. Medium dispersion gratings were used in all
cases yielding a FWHM resolution of 4-5\AA \ and a wavelength range $\lambda \lambda$
3900 - 7200 \AA . In 1992 we used EFOSC 2 at the ESO-MPI 2.2 m telescope
with the same instrumental setting as described in Motch et al. (1994).
 
All spectral and photometric data reductions were performed using standard MIDAS
procedures (Banse et al. 1983). Spectra were corrected for bias and flat-field and
later calibrated in wavelength using arc lamps. In most cases we could acquire flux
standard stars. However, uncertain weather conditions and the narrow slit entrance
width sometimes used may introduce large errors in the derived flux.

\subsection{Optical data analysis}

For active coronae, spectral classification was carried out as outlined in Motch et
al. (1997a) using Turnsheck et al. (1985), Jacoby et al. (1984) and Jaschek \& Jaschek
(1987) stellar atlases.

Visual magnitudes were in most cases not derived from our CCD imagery as they usually
lacked photometric calibration. Instead we used magnitudes extracted from the SIMBAD
database, the Guide Star Catalogue (Lasker et al. 1990) or for the faintest
counterparts, from the USNO-A1 catalogue (Monet 1997).  The GSC magnitudes of stars
having a spectral type were corrected for colour effects according to relation (1) of
Russell et al. (1990) assuming a main sequence unreddened object. After colour
correction, the remaining 1\,$\sigma$ error on magnitudes is $\approx$ 0.2 mag.  

The coordinates of the optical counterparts were in first priority extracted from the
SIMBAD database which usually gives entries from astrometric catalogues (e.g. PPM). 
When no accurate SIMBAD positions were available we used the GSC coordinates and for
the remaining identifications positions computed interactively using the Aladin sky
atlas (Bonnarel et al. 1997) at the Centre de Donn\'ees de Strasbourg (CDS).  

The Aladin project aims to provide multi wavelength cross-identification.  This tool
is designed as an interactive X-window client accessing images from the CDS image
server, Simbad database, CDS catalogue server and on-site catalogues.  The Aladin
collection contains a high resolution image archive of Schmidt plates digitized by the
Paris MAMA facility and covering a significant portion of the sky, mainly in the
Magellanic Clouds and southern Galactic Plane. The integration of the STScI Digital
Sky Survey -1 in the system provides full-sky coverage albeit with a lower spatial
resolution and astrometric accuracy than that of the MAMA archive.  For the plates
digitized by the MAMA, astrometric calibration is based on PPM standards and
reaches an accuracy better than 0.3\arcsec \ rms.

\subsection{Optical identifications}

The strategy used for optical identification of ROSAT sources in the galactic plane
has been extensively discussed in Motch et al. (1997a). For stars, we used two
criteria based on the \ion{Ca}{II} H\&K or H$\alpha$ flux to X-ray flux ratio and a
priori probability of positional coincidence in the relatively small X-ray \rqvd \ of
($<$\rqvd$>$ = 25\arcsec ). The surface density of optically bright active galactic
nuclei, cataclysmic variables, hot white dwarfs and Be stars is small enough that the
discovery of one such object in the ROSAT error circle is highly significant.  

When available, we show an optical spectrum of the identified counterpart, either low
or medium resolution but in general, do not provide finding charts as the identifier
and positions are in principle sufficient to localize the object. However in few cases
where the counterpart does not appear in the USNO catalogue we show finding charts. We
also show finding charts for all accreting sources.  

For the \nunid \ cases where we failed to find the counterpart we show on a finding
chart the observed candidates in order to ease further follow-up studies but do not
plot the spectra. 

By default, the finding charts are extracted from the STScI DSS-1 and the RGPS/SASS-I
90\% confidence error circle is shown.  In some instances where the DSS-1 data are not
able to show the candidates because of crowding or extreme colours, we use instead our
CCD images.  

Comments on individual sources, spectra and finding charts are given in section 5.

\subsection{Source selection}

The group of sources studied here was extracted from the entire RGPS source list using
four different selection criteria, hard, soft, absorbed soft and bright candidates,
all based on SASS-I hardness ratios and count rates. In this paper, we present the
optical work done on a subset of sources in each of the selected groups. Some optical
identifications which have been already published in dedicated papers are not repeated
here whereas work at the telescope is pursued for a number of other sources. To this
X-ray selected samples, we add a couple of AGN extracted from not yet fully published
lists of identifications in sample areas (e.g., GPS1-4; Motch et al.  (1991), Taurus
region; Guillout et al., (1996a)). Three other sources are from a so far barely
investigated region located at $l$ $\sim$ 130\degree.  

Hard sources were defined as having HR1$\geq$0.7 and HR2$\geq$$-$0.1 
(HR1$_{\rm BRASS}$$\geq$0.65 and HR2$_{\rm BRASS}$$\geq$$-$0.15) with the additional
condition applied in most cases that integrated galactic absorption was larger than 4
10$^{21}$ H atom cm$^{-2}$. The requirement of large galactic absorption was aimed at
screening the extragalactic component and searching preferentially for galactic
accreting binaries. Most of the sources have count rates larger than 0.1\,\cts.

Soft sources were defined as having HR1$\leq$$-$0.4 (HR1$_{\rm BRASS}$ $\leq$ $-$0.75)
with a maximal error of 0.5 and a PSPC count rate $\geq$ 0.1\,\cts. Such soft sources
do not emit much at energies higher than 0.4 keV and HR2 is therefore essentially
undefined. We expected to preferentially find white dwarfs and in general low
luminosity soft sources in this sample.

Absorbed soft sources were defined as having HR1$\geq$$-$0.4 and HR2$\leq$$-$0.4 
(HR1$_{\rm BRASS}$$\geq$$-$0.6 and HR2$_{\rm BRASS}$$\leq$$-$0.2). This sample was
designed to discover intrinsically soft luminous sources undergoing relatively large
interstellar absorption as a result of their remote location.  
  
Bright sources do not fall in any of the other hardness ratio ranges and have PSPC
count rates larger than 0.25\,\cts.

All samples were further cleaned by discarding those sources having an obvious
identification in SIMBAD (catalogued X-ray source, bright active corona). However, in
the course of this identification programme, some of these sources were recovered by
other instrumentations (Sky Lab SLX, WFC RE, EUVE, etc.) or identified by other
groups and now appear as such in SIMBAD.  

\begin{figure}
\psfig{figure=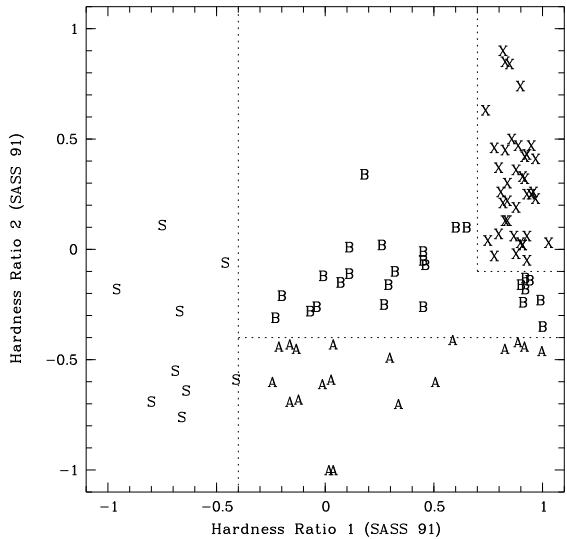,width=8.8cm,bbllx=0.0cm,bburx=22cm,bblly=0.0cm,bbury=22cm}
\caption{Position of the various X-ray selected samples in the HR1/HR2 diagram. X = 
Hard X-ray binary like, A = Absorbed supersoft candidates, S = soft white dwarf candidates,
B = remaining sources, mostly 'bright'} 
\label{selection}
\end{figure}

We show in Fig. \ref{selection} the position of the various X-ray selected samples in
the HR1/HR2 diagram and give number repartition by selection criteria in Table
\ref{xsel}.

\begin{table}
\caption{Repartition of the number of sources by selection criteria}
\begin{tabular}{lc}
\hline 
Selection & Number \\
Criteria  & of sources \\ 
\hline
Hard & 39 \\
Soft & 9\\
Absorbed soft & 19 \\
Bright & 17 \\
Other & 9\\
\hline
\end{tabular}
\label{xsel}
\end{table}

In Table 4 we list the RGPS source positions, count rates, hardness ratios and 1 RXS
identification when available.  Table 5 gives optical identification, position, B and
V when available. Spectra are shown in Fig. 8 and finding charts in Figs. 9 and 10.

\subsubsection{Hard sources}

Among the already published identifications of hard sources are the four CVs
RX\,J0028.8+5917,  RX\,J0744.9-5257, RX\,J1141.3-6410 and RX\,J2123.7+4217 (Motch et
al.  1996). Some of the Be/X-ray candidates reported in Motch et al. (1997b) or the
low-mass X-ray binary GS1826-24 identified in Barret et al. (1995) were also found in
this sample.  Here we report on the identification of RX J1739.5-2942 with a new
Be/X-ray binary (see below). Our hard sample also contains the ultrasoft transient SLX
1746-331 (Skinner et al. 1990) which was apparently in outburst at the time of the
ROSAT survey observation in September 1990 and not detected in follow-up ROSAT HRI
observations on 1994 October 2.

\begin{figure}
\psfig{figure=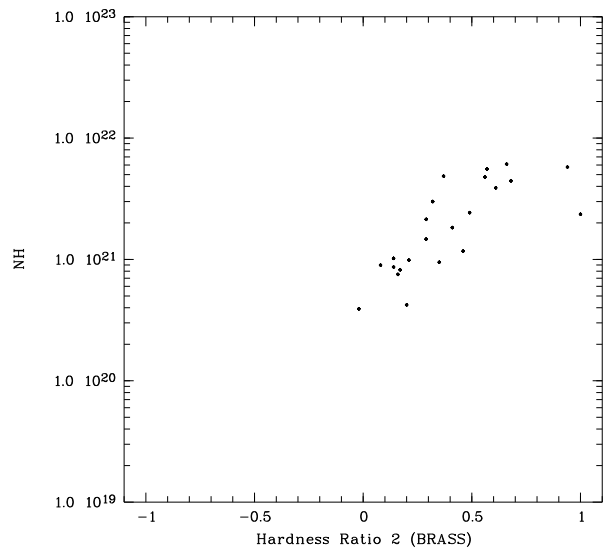,width=8.8cm,bbllx=0.0cm,bburx=22cm,bblly=0.0cm,bbury=22cm}
\caption{Relation between hardness HR2 and galactic absorption for the newly identified
AGN}
\label{agn_id_h2}
\end{figure}

SIMBAD contains hardly any heavily absorbed AGN whereas all luminous galactic X-ray
binaries, mostly discovered at higher energies are listed. This strong bias indicates
that in hard X-ray selected samples, the number ratio of AGN to X-ray binaries or CVs
should be much larger than suggested in Fig. \ref{all_h1h2}, \ref{hr1_nh} and
\ref{hr2_nh}.  A total of 13 AGN are indeed found in the hard sample, confirming that
in the galactic plane, there is no easy way to disentangle absorbed AGN from genuine
accreting binaries since both populations exhibit hard spectra and faint bluish
counterparts.  Not surprisingly, the new identified hard AGN are much more absorbed
than the SIMBAD sample and their HR2 is correlated with NH (see Fig. \ref{agn_id_h2}).
In 10 instances we identify the X-ray source with an optically bright active corona
and in 13 cases, we fail to find a likely counterpart. Not unexpectedly, the vast
majority of unidentified sources (13 among 17) are found in the hard sample and based
on the properties of the identified population we can predict that most of these sources
are likely to be absorbed extragalactic objects. In particular, from the lack of
bright objects in I band images we can exclude relatively close Be/X-ray binaries as
possible identification for these hard sources.

\subsubsection{Soft sources}

The majority of the very soft sources discovered in the RASS turned out to be white
dwarfs often also detected as bright UV sources in the Wide Field Camera Survey
(Pounds et al.  1993) or by the Extreme Ultraviolet Explorer (Malina et al. 1994). 
Many of these bright sources were readily optically identified from their UV positions
(Mason et al.  1995).  Again, some active coronae emit low temperature X-ray spectra
which are indistinguishable from white dwarf emission at the PSPC survey sensitivity.

Among the already published discoveries in this soft sample are the extremely hot PG
1159 star RX J2117.1+3412 (Motch et al.  1993) and the hot WD RX J2052.7+4639 (Motch
et al. 1997a).

The other class appearing in this X-ray selected group is that of soft polars and
intermediate polars in which the blackbody like X-ray emission from the polar cap
dominates over the hard bremsstrahlung radiation from the shock in the accretion
column (Beuermann \& Schwope 1994, Haberl \& Motch 1995). The polar RX J1802.1+1804
(Greiner et al. 1995) and RE 0751+14 (Mason et al. 1992, Motch \& Haberl 1995) were
also found in this sample. The exotic population of isolated neutron stars accreting
from the interstellar medium should also show up as soft sources. In fact, two of the
best candidates known so far were found in the soft sample (RXJ1856-37 Walter et al. 
1996, RX J0720-3125 Haberl et al. 1997). RX J2102.0+3359 (see Tables 4 and 5) was also
considered for some time as a good lone neutron star candidate.  However, UB CCD
imagery revealed the presence of a faint UV excess object (B= 21.6, U$-$B = $-$1.1)
close to the center of the error circle. This picture hints towards identification
with an extreme \fxfo \ ratio AM Her CV such as RXJ0453.4-4213 = RS Cae (Burwitz et
al. 1996).  However, its firm identification awaits optical spectroscopic confirmation.

In the soft sample presented here, we report the identification of a new DA white
dwarf (RX J1936.3+2632 = 2EUVE J1936+26.5), 2 new polars (RX J0649.8-0737 and RX
J0749.1-0549) of rather long orbital period and of a couple of active coronae.  

\subsubsection{Absorbed soft sources}

Based on the obvious observation that no extremely X-ray bright soft source was found
in the ROSAT survey nor in previous EUV surveys, we concluded that if a luminous
supersoft source analogue to Cal 83 in the LMC were existing in the Galaxy, its
observed X-ray spectrum would be dimmed and distorted by interstellar absorption. In
order to delimit the range of hardness values to select, we simulated soft black body
spectra affected by large interstellar absorption. With the ROSAT PSPC, large
photoelectric absorption of a very soft source produces a very peaked distribution
centered at an energy which depends on the balance between kT and \nh .  

Unfortunately, at the PSPC resolution, optically thin thermal spectra with many
emission lines in the range of 0.3-0.5 keV can also produce peaked spectra, albeit
with a much lower photoelectric absorption. Some active coronae may emit this kind of
spectra in a narrow range of temperature and photoelectric absorption. Such spectra
are also often seen in supernova remnants and in the hot diffuse galactic emission in
general. Care was therefore taken to exclude extended X-ray sources by examining the
survey images.  

The first sample of absorbed soft sources in which we discovered the galactic
supersoft binary RX J0925-4758 (Motch et al. 1994) and the peculiar intermediate
polar candidate RX J1914.4+2456 (Motch et al. 1996) had more stringent constraints
than set here (HR1$_{\rm SASS-I}$ $\geq$ $-$0.4 and HR2$_{\rm SASS-I}$ $\geq$ $-$0.6
with errors on HR1 and HR2 of less than 0.5 and 1.0 respectively and a
count rate larger than 0.1\,\cts). This restricted sample is now mostly identified and
the remaining sources are associated with active stars.

Relaxing the constraints on HR2 increased the number of candidates but also the
fraction of active coronae. Therefore, apart from one AGN (RX J1929.8+4622) and two
CVs (RX J1951.7+3716 and RX J1946.2-0444), most sources are identified with stars. In
three cases, however, we fail to identify the source. A fraction of these unidentified
sources could be AGN with absorbed steep spectra.

\subsubsection{Bright sources}

A number of unidentified bright RGPS sources of high \fxfo \ turned out to be
previously unknown AGN or CVs. As shown in Fig. \ref{ratio_12}, selecting high \fxfo \
sources at intermediate hardness ratios does improve the fraction of extragalactic and
cataclysmic variable identifications.  Some very active coronae with magnitudes below
that of the limit of HD or SAO catalogues were also discovered.  

\section{Conclusions}

The BRASS/SIMBAD cross-correlation and the set of optical identifications presented in
this paper show that careful selection on multi-wavelength parameters can help to
disentangle the various populations of X-ray sources accounting for the ROSAT survey.

In order to reach the state at which one can really compute probabilities of
identification of any source with the main classes of X-ray emitters, large source
samples have to be completely identified using catalogue data and proper
multi-wavelength observations.  In the galactic plane, the level of screening of the
extragalactic population and in general galactic latitude are key parameters. Ideally,
selected test samples should be accumulated at various latitudes in order to properly
handle likelihoods of identification such as for instance outlined in Sterzik et al. 
(1995).

Among the \ns \ selected sources, we find a likely counterpart to the X-ray source in
\nid \  cases.  With 39 identifications, active coronae again dominate source count.
However, X-ray selection enhances the number of non-coronal identifications and we
report on 25 new AGN, 6 new CVs, 2 new white dwarfs and a new Be/X-ray binary.

Our programme of optical identification of X-ray selected RGPS sources is going on and
new identifications will be reported in a forthcoming paper. Statistics of
identifications in the various selected samples will also be presented at this
occasion.

\section{Notes on individual objects}

\subsection{Stars}

 \noindent \newline RX J0222.5+5033 = BD+49 646 :
BD+49 646 was also detected by the ROSAT WFC (2RE J0222+503, Pye et al. 1995). 
The star GSC0330200570 located NE from BD+49 646 is a late A type star unrelated to the X-ray source. 
 \noindent \newline RX J0621.2+4415 = G 101-35 :
The BRASS error circle now encompasses the Me star. The star has high proper motion.
 \noindent \newline RX J0635.9+0755 = GSC0073302098 :
The candidate star GSC0073302098 lacks the strong \ion{Ca}{II} H\&K emission which
would be expected based on the PSPC count rate. However, the a priori probability 
of coincidence of the ROSAT source with the late G-K V $\approx$ 10.7 star is small
enough so that the identification can be considered. The ROSAT source is identified
 with 2E 0633.2+075. \noindent \newline RX J0702.0+1257 = GSC0075701608 :
Using EUVE data, Vennes et al. (1997) identify the source as a K0IV-V and a hot white
dwarf in a wide binary. Part of the softness of the PSPC spectrum could be indeed
explained by a contribution from the degenerate star.  The XRT source is identical with
EUVE J0702+129 and 2RE J0702+125 (Pye et al. 1995).
 \noindent \newline RX J0704.5-0612 = GSC0482603053 ?? :
The brightest object in the RGPS error circle is GSC0482603053. A medium resolution
spectrum of the star reveals \ion{Ca}{II} H\&K emission and some evidence for weak \Halp
\ emission. Unfortunately, no absolute CaII flux could be measured for this active
corona, leaving some doubt on the compatibility of the chromospheric signature with the
ROSAT X-ray flux. The second brightest object located SW from GSC0482603053 is a late
type star without marked activity. 
 \noindent \newline RX J0713.1-0511 = GSC0482302265 :
The XRT source is identical to 2RE J0713-051. 
 \noindent \newline RX J0721.3-5720 = GSC0855901016 + comp :
The candidate M5eV star exhibits strong Balmer emission and is probably responsible for a
large part of the X-ray emission. The close and comparatively bright star (V$\approx$ 11)
GSC0855901016 is a main sequence K0 star. Unfortunately, we do not have \ion{Ca}{II} H\&K
spectra which would have given indication on the level of stellar activity. The a priori
coincidence of position between GSC0855901016 and a ROSAT survey source is small and
could indicate that the bright star also contributes to the X-ray emission and
constitutes a physical pair with the Me star.   \noindent \newline RX J0811.1-5555 = GSC0857001980    :
The absence of a spectrum covering the \ion{Ca}{II} H\& K region does not allow to firmly
identify this source. However, in this case, the BRASS position which has a better accuracy than
the RGPS determination, is centered on GSC0857001980 and excludes any other object than
the V $\approx$ 12 late G star. This positional evidence and the soft X-ray hardness
ratios suggest that GSC0857001980 is the counterpart of the ROSAT source.
 \noindent \newline RX J0828.5-5138 = GSC0816200330    :
The BRASS position is now centered on the G star GSC0816200330. The low a priori
probability of a random positional coincidence of the X-ray source with the relatively
bright (V $\approx$ 11) star and the soft X-ray hardness ratio suggest a
coronal identification.   \noindent \newline RX J0845.7-3544 = GSC0714900583    :
GSC0714900583 is a late G star, which considering the coincidence in position with the X-ray source and the soft X-ray spectrum is most probably the optical counterpart of the X-ray source.
 \noindent \newline RX J0856.4-2241 = GSC0658500334 :
Weak Balmer emission may be present in the spectrum of GSC0658500334. The BRASS position
is also centered on the early M star with a similar error radius. The soft X-ray spectrum
is also consistent with a coronal identification. \noindent \newline RX J1704.3-4020 = HD 322763  :
A follow-up HRI observation gives an improved position
17H04M17.68S $-$40D19M47.6S centered on the A star.
 \noindent \newline RX J2010.5+0632 = GSC0050701588 :
The Me star is probably identified with the Parkes-MIT-NRAO survey source PMN
J2010+0632 (Griffith et al. 1995). An X-ray flare was detected during the survey.
 \noindent \newline RX J2014.8+4501 = HD 192785 :
The K0V star HD 192785 exhibits \Halp \ reemission (not shown) and was also detected
in the Einstein Slew Survey (Schachter et al. 1996).

\subsection{Active galactic nuclei}

AGN identified in the low galactic latitude regions have all low redshift
and the dominant population is Seyfert 1. 

\begin{table}
\caption{Magnitude and redshift of newly identified AGN}
\label{zagn}

\begin{tabular}{cclc}
Source Name    &    V/B & Type  & $z$ \\
\hline
 RX J0222.1+5221 & 17.00 &Sy 1     &    0.200 \\
 RX J0254.6+3931 & 16.10 &BLRG      &   0.289\\
 RX J0324.7+3410 & 15.10 &Sy 1      &   0.063\\
 RX J0325.2+4042 & 15.20 &Sy 2:     &   0.048\\
 RX J0337.0+4738 & 17.00 &Sy 1      &   0.184\\
 RX J0434.7+4014 & 15.20 &Sy 1      &   0.021\\
 RX J0452.0+4932 & 17.10 &Sy 1      &   0.029\\
 RX J0459.8+1808 & 19.50 &Sy 1      &   0.157\\
 RX J0508.3+1721 & 13.47 &Sy 2:     &   0.017\\
 RX J0508.9+2113 & 17.70 &Sy 1      &   0.190\\
 RX J0602.1+2828 & 15.00 &Sy 1      &   0.033\\
 RX J0608.0+3058 & 17.30 &Sy 1      &   0.073\\
 RX J0750.9+0320 & 15.20 &Sy 1      &   0.099\\
 RX J0755.7$-$0157 & 14.80 &Sy 1    &     0.040\\
 RX J0801.9$-$4946 & 11.07 &Sy 1 &     0.040\\
 RX J0816.4$-$1311 & 17.40 &BL LAC  &          *\\
 RX J0818.9$-$2252 & 15.50 &Sy 1    &     0.035\\
 RX J0913.0$-$2103 & 17.30 &BL LAC  &          *\\
 RX J1023.9$-$4336 & 17.50 &BL LAC  &          *\\
 RX J1741.4+0348 & 15.30 &Sy 1      &   0.030\\
 RX J1929.8+4622 & 17.00 &Sy 1      &   0.127\\
 RX J1931.1+0937 & 19.00 & BL LAC   &         *\\
 RX J2040.3+1059 & 16.10 &Sy 1      &   0.085\\
 RX J2043.9+5314 & 17.50 &Sy 1      &   0.080\\
 RX J2044.0+2833 & 14.70 &Sy 1      &   0.050\\
\hline
\end{tabular}
\end{table}

 \noindent \newline RX J0222.1+5221 = PMM1350-02324516 :
This identification was alredy reported in Motch et al. (1991). \noindent \newline RX J0324.7+3410 = GSC0234901904 :
This Seyfert 1 nucleus was independently discovered and identified with the HEAO-1 source
H0321+340 by Remillard et al. (1993) and with the ROSAT source by Kock et al. (1996). \noindent \newline RX J0452.0+4932 = H0432052 :
This object was already listed by Hauschildt (1987). \noindent \newline RX J1931.1+0937 = A  :
Follow up ROSAT HRI observations revealed flare like activity by a factor of
$\approx$ 2 on a time scale of one day. The best HRI position is 19H31M09.23S
+09D37M18.96S with a conservative error radius of 10\arcsec. 
The proposed optical counterpart displays a blue featureless continuum and is located 
within the 1\arcsec \ radius error circle of the radio source NVSS J193109+093717. Based on
these evidences we propose an identification with a BL Lac type of AGN. 
 \noindent \newline RX J2040.3+1059 = PMM0975-19711620 :
The AGN could be in an interacting system

\subsection{Accreting sources}

Detailed X-ray and, optical photometric and spectroscopic studies of the new CVs will
be presented in future papers.  

 \noindent \newline RX J0649.8-0737 = PMM0750-02887019 :
This CV is a polar with a $\approx$ 4.4h orbital period. \noindent \newline RX J0749.1-0549 = PMM0825-05603282 :
This CV is a polar with a $\approx$ 3.6h orbital period. \noindent \newline RX J1739.4-2942 = A :
RX J1739.5-2942 is probably identical with GRS 1736-197 since the ROSAT position is
located well within the 90\arcsec \ radius of the 90\% confidence circle of GRS 1736-197.
The Be/X-ray nature of the source is consistent with the hard ART-P X-ray spectrum
observed by GRANAT (Pavlinsky et al. 1994). 

\subsection{Miscellaneous and optically unidentified}
         
 \noindent \newline RX J0035.8+5950 = ?? :
None of the two GSC stars GSC0366601407 (A) and GSC0366600907 (B) lying at the edge of the
ROSAT error circle exhibits detectable chromospheric activity.
 \noindent \newline RX J0529.0+0934 = ?? :
Objects A and B display featureless continuum. B is bluer than A.  
 \noindent \newline RX J0602.2+2837 = ?? :
Object A has a featureless continuum. Object B could exhibit H$\alpha$ emission.
 \noindent \newline RX J0620.6+2644 = ?? :
The BRASS 90\% confidence error area (small circle) does not overlap with the RGPS 90\%
error area. Because of the improved reduction process, the BRASS position should be
considered as the best one. All objects A-E are late type stars without Balmer emission.
A: late K, B: G-K, C: K, D: G-K and E: late G.
 \noindent \newline RX J0621.7+1747 = ?? :
Object A is a late B type star, B a K type star and C a G-K type star.
 \noindent \newline RX J0648.7+1516 = ??                   :
The ROSAT source is identical to 1H 0646+152. A short exposure time spectrum of object A 
reveals a rather blue featureless continuum. B and C are probably late type stars.
 \noindent \newline RX J0717.4-1119 = ?? :
The source was not recovered in a subsequent HRI pointing. Object A is a K type star and
B has an A spectral type.
 \noindent \newline RX J0759.1+0748 = ?? :
Bright X-ray source. Object A has an F-G spectral type and object B displays
\Halp \ and \Hbet \ in absorption. Objects B and C have reddish featureless continua.
 \noindent \newline RX J0819.2-0756 = ?? :
Objects A and B are F-G type stars. Object C is the only object visible inside the
HRI position. \noindent \newline RX J1718.4-4029 = ?? :
The image shown here is a 5 min I band CCD exposure obtained in April 1992 with EFOSC2 on
ESO-MPI 2.2m. The HRI error circle (RA = 17H18M24.13S, $-$40D29M30.4S,
 90\% confidence radius = 15.7\arcsec, 2000.0 eq) is plotted. Objects A,B and C
display \Halp \ and \Hbet \ absorption lines.  
 \noindent \newline RX J1740.7-2818 = ?? :
The CCD image shown here is a 5 min I band exposure obtained with EFOSC2 and the 
MPI-ESO 2.2m telescope in April 1992. The field is heavily reddened. Objects A,B,C and D
are late M stars without any noticeable Balmer emission. \noindent \newline RX J1742.3-2737 = ?? :
The CCD image shown here is a 2 min I band exposure obtained with EFOSC2 and the 
MPI-ESO 2.2m telescope in April 1992. The field is heavily reddened. Object A is a
reddened A-F type star and object B a reddened G-K star. \noindent \newline RX J1749.8-3312 = SLX 1746-331 :
The CCD image shown here is a 5 min I band exposure obtained with EFOSC2 and the MPI-ESO
2.2m telescope in April 1992. The field is heavily reddened. Object A is a G-K type star.
RX J1749.8-3312 = 1RXS J174948.4-331215 is probably identified with the soft transient
SLX 1746-331 (Skinner et al. 1990). The source was discovered by the Spacelab-2
coded-mask X-ray telescope in July-August 1985. Based on its ultrasoft spectrum and
transient nature White \& van Paradijs (1996) classify the source as a black hole
transient. During the ROSAT survey observation (1990 Sep 08 23:15:57 to 1990 Sep 10
07:18:27) the source was caught in outburst and not detected during a subsequent ROSAT
HRI follow-up observation (1994 Oct 2 from 05:58:58 to 06:30:40 UT). Considering
the expected low mass of the companion star and high interstellar absorption on the line
of sight, the optical identification is likely to be difficult outside X-ray outbursts.   
\noindent \newline RX J1804.1+0042 = PMM0900-11515260 :
Follow-up ROSAT HRI observations show the source to be extended. Its mean position is
18H04M08.6S  +00D42M26.3S, well centered on an extended optical object which could be an
elliptical galaxy. A low S/N spectrum suggests a possible redshift of $z$ = 0.07 $\pm$
0.02. The galaxy is probably identified with the Parkes-MIT-NRAO radio source PMN
J1804+0042 (Brinkmann et al. 1997). \noindent \newline RX J1936.3+2632 = A  :
= 2EUVE J1936+26.5
 \noindent \newline RX J1943.9+2118 = ??  :
We show a 10 min I band CCD exposure obtained in September 1995 with the 1.2\,m telescope
at OHP. The large error circle represents the RGPS localisation whereas the small one is
derived from a follow-up ROSAT HRI observation. The HRI position is 19H43M56.16S
+21D18M24.8S with a conservative error circle of 10\arcsec . A is a reddened
object without marked features apart from the Na I line and a possible Mg G band hollow.
The position of A, 19H43M56.2S +21D18M20.4S is outside the error circle of the radio
source NVSS J194356+211826 which is itself close to the center of the HRI error circle.
The overall radio, optical and X-ray picture favors an heavily absorbed extragalactic
identification. 
 \noindent \newline RX J1950.0+3821 = ?? :
Objects A, B, C and D are G type stars. The RGPS error circle contains a radio source
NVSS J195004+382210 which is located $\approx$ 8\arcsec \ north of A.
 \noindent \newline RX J2102.0+3359 = ?? UV excess object :
We show here the BRASS position overlayed on a 10 min B filter CCD image obtained with
the OHP 1.2\,m telescope in September 1995. Object A is a late G early K star, B a G star. 
A low S/N spectrum reveals a featureless continuum for object C. Object X,
located at 21H02M01.4S +33D59M29.6S exhibits a strong UV excess with B = 21.6 and U-B =
$-$1.1$\pm$ 0.3. The softness of the X-ray source and the presence of an UV excess object
in the ROSAT error circle argue in favour of a AM Her system. Final identification should
however await spectroscopic observations.  
 \noindent \newline RX J2104.2+2118 = ?? :
Objects A and B are late type stars. Object E displays a blue featureless continuum. 
The radio source NVSSJ210415+211805, is located in the RGPS and BRASS error circles,
$\approx$ 2\arcsec \ southern of object F. 
 \noindent \newline RX J2156.3+3318 = PMM1200-18997159 :
X-ray emission is slightly extended by 1 arcmin and centered on the presumably elliptical
galaxy PMM1200-18997159 ($z$ = 0.079, B=16.2, R=12.4).  The fx/fopt ratio is compatible
with normal emission from the galaxy without the need to add cluster contribution. The
radio source NVSS J215623+331837 is coincident with the galaxy.  

\begin{acknowledgements}

We thank the night assistants at Observatoire de Haute-Provence for carrying out some
of the observations at the 1.2\,m telescope. The ROSAT project is supported by the
Bundesministerium f\"ur Bildung, Wissenschaft, Forschung und Technologie (BMBF/DLR)
and the Max-Planck-Gesellschaft.  C.M.  acknowledges support from a CNRS-MPG
cooperation contract and thanks Prof. J. Tr\"umper and the ROSAT group for their
hospitality and fruitful discussions. We are particularly grateful to the MPE team
for providing early access to RASS data. This research has made use of the
ALADIN sky atlas and of the SIMBAD database operated at CDS, Strasbourg, France. We
thank  F.  Ochsenbein for providing easy access to the GSC and to the USNO-A1.0
catalogues.  Finding charts were extracted from the \it Digitized Sky Survey, \rm
produced at the Space Telescope Science Institute (ST ScI) under U.S. Government grant
NAG W-2166.

\end{acknowledgements}

\onecolumn

\begin{table*}

   \caption[]{\label{tab_rgps} X-ray characteristics of ROSAT survey sources
derived from the SASS as in October 1991. Coordinates are equinox 2000.0. The second
column indicate the selection origin of the source as shown on Fig. 6, S, soft, A,
absorbed soft, X, X-ray binary like, i.e. hard, B, remaining sources, mostly bright.
Whenever the SASS-I RGPS source is bright enough to appear in the BRASS catalogue we
give the 1RXS name}

\end{table*}
\scriptsize
\tabcolsep=5pt
\tablehead{\hline \tbsp
Source & Sel & Right      & Declination & \rqvd    &Cnt rate  & Error  & HR1 & Error
& HR2 & Error & BRASS  \\[0.8ex]
Name   &       & Ascension  &             & \arcsec& \cts  & \cts&     &
&     &        & Name \\[0.8ex]
\hline \tbsp}
\tabletail{\hline \multicolumn{10}{l}{\sl Continued on next page}\\ \hline}
\tablelasttail{\hline}
\begin{supertabular}{crcrcccrrrrl}

RX J0035.8+5950 &X&00H35M52.6S & 59D50M08S  &19.2&0.337&0.037& 0.91&0.06& 0.33&0.08&1RXS J003552.8+595006\\   
RX J0119.0+7033 &X&01H19M02.8S & 70D33M09S  &25.9&0.145&0.020& 0.93&0.07& 0.43&0.14&1RXS J011859.8+703336\\   
RX J0143.7+6349 &A&01H43M44.0S & 63D49M30S  &21.5&0.051&0.013& 0.02&0.53&$-$1.00&9.99&1RXS J014343.7+634931\\   
RX J0146.0+6348 &A&01H46M05.8S & 63D48M52S  &21.5&0.054&0.017& 0.04&0.71&$-$1.00&9.99&1RXS J014605.5+634853\\   
RX J0202.2+6934 &B&02H02M14.7S & 69D34M23S  &21.5&0.042&0.010& 0.07&0.52&$-$0.15&0.52&                     \\   
RX J0222.1+5221 &X&02H22M06.9S & 52D21M11S  &25.9&0.174&0.028& 0.86&0.13& 0.50&0.15&1RXS J022206.0+522112\\   
RX J0222.5+5033 &B&02H22M35.0S & 50D33M34S  &18.4&0.422&0.034&$-$0.07&0.08&$-$0.28&0.11&1RXS J022234.1+503335\\   
RX J0254.6+3931 &B&02H54M41.8S & 39D31M41S  &20.3&0.121&0.020& 1.00&9.99&$-$0.35&0.16&1RXS J025441.4+393143\\   
RX J0256.3+6141 &A&02H56M20.7S & 61D41M24S  &21.5&0.153&0.021&$-$0.01&0.14&$-$0.61&0.32&1RXS J025620.0+614129\\   
RX J0324.7+3410 &X&03H24M42.2S & 34D10M51S  &21.5&0.235&0.026& 0.91&0.08& 0.02&0.11&1RXS J032441.3+341056\\   
RX J0325.2+4042 &B&03H25M17.6S & 40D42M00S  &27.5&0.045&0.013& 0.45&0.42&$-$0.01&0.53&                     \\   
RX J0337.0+4738 &X&03H37M04.0S & 47D38M53S  &21.5&0.134&0.021& 0.83&0.16& 0.13&0.17&1RXS J033703.9+473852\\   
RX J0416.0+5237 &A&04H16M03.0S & 52D37M31S  &22.9&0.102&0.025&$-$0.24&0.28&$-$0.60&0.45&1RXS J041602.7+523737\\   
RX J0434.7+4014 &X&04H34M42.2S & 40D14M21S  &19.2&0.197&0.024& 0.84&0.15& 0.22&0.13&1RXS J043442.1+401425\\   
RX J0452.0+4932 &X&04H52M05.0S & 49D32M46S  &17.7&0.595&0.048& 0.92&0.07& 0.42&0.06&1RXS J045205.0+493248\\   
RX J0459.8+1808 &B&04H59M51.2S & 18D08M47S  &38.5&0.061&0.014& 0.60&0.33& 0.10&0.26&                     \\   
RX J0508.3+1721 &X&05H08M20.6S & 17D21M38S  &22.9&0.064&0.018& 0.89&0.10& 0.47&0.40&1RXS J050820.9+172134\\   
RX J0508.9+2113 &X&05H08M55.7S & 21D13M04S  &31.0&0.058&0.018& 0.88&0.12&$-$0.02&0.18&1RXS J050855.3+211304\\   
RX J0529.0+0934 &X&05H29M03.1S & 09D34M38S  &18.4&0.431&0.042& 0.96&0.04& 0.26&0.07&1RXS J052902.7+093439\\   
RX J0535.0+6450 &S&05H35M00.7S & 64D50M47S  &25.9&0.160&0.034&$-$0.75&0.22& 0.11&0.70&1RXS J053501.1+645046\\   
RX J0554.7+1055 &A&05H54M45.8S & 10D55M59S  &21.5&0.193&0.033&$-$0.16&0.12&$-$0.69&0.27&1RXS J055446.0+105559\\   
RX J0602.1+2828 &X&06H02M10.6S & 28D28M25S  &18.4&0.708&0.057& 0.97&0.03& 0.23&0.06&1RXS J060210.7+282821\\   
RX J0602.2+2837 &X&06H02M17.5S & 28D37M04S  &25.9&0.149&0.031& 0.93&0.06&$-$0.05&0.13&                     \\   
RX J0608.0+3058 &X&06H08M01.7S & 30D58M53S  &22.9&0.112&0.025& 0.78&0.19& 0.46&0.40&1RXS J060801.7+305847\\   
RX J0620.6+2644 &X&06H20M40.3S & 26D44M28S  &25.9&0.338&0.038& 0.93&0.07& 0.06&0.13&1RXS J062040.0+264339\\   
RX J0621.2+4415 &A&06H21M13.8S & 44D15M47S  &21.5&0.510&0.074& 0.04&0.14&$-$0.43&0.19&1RXS J062113.1+441430\\   
RX J0621.7+1747 &X&06H21M47.8S & 17D47M30S  &19.2&0.113&0.021& 0.84&0.15& 0.30&0.46&1RXS J062148.1+174736\\   
RX J0625.8$-$0101 &S&06H25M49.8S &$-$01D01M40S  &21.5&0.143&0.022&$-$0.46&0.16&$-$0.06&0.53&1RXS J062549.8$-$010138\\   
RX J0635.9+0755 &X&06H35M58.2S & 07D55M28S  &20.3&0.247&0.029& 0.81&0.17& 0.26&0.11&1RXS J063558.5+075527\\   
RX J0648.7+1516 &X&06H48M47.4S & 15D16M26S  &   *&0.679&0.679& 1.03&0.06& 0.03&0.06&1RXS J064847.8+151626\\   
RX J0649.8$-$0737 &S&06H49M50.2S &$-$07D37M41S  &21.5&0.557&0.055&$-$0.69&0.08&$-$0.55&0.36&1RXS J064949.8$-$073734\\   
RX J0656.8$-$1424 &X&06H56M53.6S &$-$14D24M56S  &17.7&0.788&0.048& 0.80&0.05& 0.07&0.06&1RXS J065653.2$-$142455\\   
RX J0702.0+1257 &S&07H02M04.1S & 12D57M50S  &18.4&1.180&0.067&$-$0.67&0.04&$-$0.28&0.15&1RXS J070204.3+125758\\   
RX J0704.5$-$0612 &X&07H04M31.3S &$-$06D12M35S  &24.3&0.113&0.021& 0.78&0.20&$-$0.03&0.18&1RXS J070431.1$-$061224\\   
RX J0713.1$-$0511 &B&07H13M11.1S &$-$05D11M52S  &17.7&1.470&0.069&$-$0.23&0.05&$-$0.31&0.07&1RXS J071311.0$-$051142\\   
RX J0717.4$-$1119 &X&07H17M25.1S &$-$11D19M43S  &27.5&0.145&0.026& 0.83&0.16& 0.45&0.13&1RXS J071722.6$-$111954\\   
RX J0721.3$-$5720 &B&07H21M22.8S &$-$57D20M49S  &32.8&0.199&0.020&$-$0.01&0.10&$-$0.12&0.13&1RXS J072123.9$-$572034\\   
RX J0749.1$-$0549 &S&07H49M09.7S &$-$05D49M31S  &27.5&0.615&0.082&$-$0.80&0.18&$-$0.69&0.37&1RXS J074909.8$-$054933\\   
RX J0750.9+0320 &B&07H50M59.8S & 03D20M18S  &18.4&0.808&0.057& 0.32&0.07&$-$0.10&0.09&1RXS J075100.0+032017\\   
RX J0755.7$-$0157 &B&07H55M47.3S &$-$01D57M48S  &17.7&0.882&0.058& 0.29&0.07&$-$0.16&0.09&1RXS J075547.4$-$015742\\   
RX J0759.1+0748 &A&07H59M09.0S & 07D48M26S  &21.5&3.360&0.200&$-$0.16&0.06&$-$0.43&0.09&1RXS J075908.8+074835\\   
RX J0801.9$-$4946 &X&08H01M57.8S &$-$49D46M42S  &20.3&0.173&0.018& 0.92&0.07& 0.32&0.10&1RXS J080157.7$-$494639\\   
RX J0811.1$-$5555 &B&08H11M09.8S &$-$55D55M26S  &36.5&0.125&0.014& 0.18&0.11& 0.34&0.16&1RXS J081108.7$-$555553\\   
RX J0816.4$-$1311 &B&08H16M26.7S &$-$13D11M46S  &19.2&0.861&0.058& 0.92&0.03&$-$0.13&0.08&1RXS J081626.9$-$131149\\   
RX J0818.9$-$2252 &X&08H18M57.3S &$-$22D52M32S  &24.3&0.566&0.047& 0.90&0.07& 0.03&0.08&1RXS J081858.0$-$225229\\   
RX J0819.2$-$0756 &A&08H19M17.7S &$-$07D56M20S  &20.3&0.158&0.029& 0.83&0.16&$-$0.45&0.18&1RXS J081917.6$-$075620\\   
RX J0828.5$-$5138 &B&08H28M33.2S &$-$51D38M19S  &25.9&0.236&0.021& 0.11&0.10&$-$0.11&0.11&1RXS J082832.2$-$513828\\   
RX J0845.7$-$3544 &A&08H45M43.0S &$-$35D44M27S  &22.9&0.077&0.016& 0.34&0.46&$-$0.70&0.26&                     \\   
RX J0856.4$-$2241 &A&08H56M26.4S &$-$22D41M42S  &25.9&0.102&0.025& 0.30&0.48&$-$0.49&0.40&1RXS J085626.3$-$224141\\   
RX J0913.0$-$2103 &B&09H13M00.8S &$-$21D03M23S  &17.7&0.924&0.052& 0.94&0.02&$-$0.14&0.07&1RXS J091300.4$-$210315\\   
RX J0935.5$-$2802 &B&09H35M31.5S &$-$28D02M50S  &21.5&0.267&0.031&$-$0.04&0.11&$-$0.26&0.17&1RXS J093530.9$-$280255\\   
RX J1023.9$-$4336 &B&10H23M55.5S &$-$43D36M03S  &19.2&1.280&0.062& 0.92&0.03&$-$0.18&0.05&1RXS J102356.4$-$433602\\   
RX J1155.4$-$5641 &B&11H55M27.7S &$-$56D41M59S  &31.0&0.819&0.067& 0.65&0.05& 0.10&0.06&1RXS J115527.6$-$564149\\   
RX J1606.6$-$4618 &X&16H06M37.1S &$-$46D18M30S  &31.0&0.139&0.025& 0.84&0.15& 0.13&0.17&1RXS J160637.1$-$461835\\   
RX J1639.7$-$3920 &X&16H39M47.8S &$-$39D20M15S  &31.0&0.254&0.033& 0.80&0.18& 0.37&0.13&1RXS J163947.8$-$392023\\   
RX J1704.3$-$4020 &X&17H04M19.4S &$-$40D20M13S  &21.5&0.214&0.039& 0.88&0.12& 0.36&0.13&1RXS J170419.7$-$402012\\   
RX J1718.4$-$4029 &X&17H18M24.2S &$-$40D29M34S  &31.0&0.167&0.039& 0.74&0.23& 0.63&0.31&1RXS J171824.2$-$402934\\   
RX J1739.4$-$2942 &X&17H39M30.1S &$-$29D42M07S  &19.2&0.154&0.030& 0.85&0.14& 0.84&0.15&1RXS J173930.3$-$294211\\   
RX J1740.7$-$2818 &X&17H40M42.8S &$-$28D18M03S  &20.3&0.336&0.041& 0.83&0.15& 0.85&0.14&1RXS J174043.1$-$281806\\   
RX J1741.4+0348 &X&17H41M27.5S & 03D48M48S  &18.4&1.180&0.063& 0.91&0.09& 0.02&0.05&1RXS J174128.1+034848\\   
RX J1742.3$-$2737 &X&17H42M20.2S &$-$27D37M36S  &24.3&0.166&0.033& 0.90&0.10& 0.74&0.22&1RXS J174220.8$-$273736\\   
RX J1749.8$-$3312 &X&17H49M48.3S &$-$33D12M26S  &18.4&0.543&0.062& 0.97&0.03& 0.41&0.07&1RXS J174948.4$-$331215\\   
RX J1804.1+0042 &X&18H04M08.3S & 00D42M29S  &22.9&0.354&0.042& 0.95&0.05& 0.25&0.09&1RXS J180408.7+004229\\   
RX J1925.0+4429 &A&19H25M01.8S & 44D29M43S  &21.5&0.317&0.030& 0.51&0.08&$-$0.60&0.08&1RXS J192502.2+442948\\   
RX J1929.8+4622 &A&19H29M50.1S & 46D22M16S  &21.5&0.392&0.030& 0.59&0.09&$-$0.41&0.07&1RXS J192949.7+462231\\   
RX J1931.1+0937 &X&19H31M09.0S & 09D37M22S  &17.7&0.625&0.047& 0.93&0.07& 0.25&0.08&1RXS J193109.5+093714\\   
RX J1935.4+3746 &A&19H35M29.4S & 37D46M09S  &25.9&0.528&0.042&$-$0.13&0.08&$-$0.45&0.12&1RXS J193528.9+374605\\   
RX J1936.3+2632 &S&19H36M18.1S & 26D32M42S  &31.0&0.100&0.022&$-$0.64&0.30&$-$0.64&0.42&                     \\   
RX J1943.9+2118 &X&19H43M55.4S & 21D18M15S  &32.8&0.103&0.023& 0.82&0.17& 0.90&0.09&1RXS J194356.1+211731\\   
RX J1946.2$-$0444 &A&19H46M16.5S &$-$04D44M47S  &21.5&0.114&0.026& 1.00&9.99&$-$0.46&0.18&1RXS J194616.9$-$044456\\   
RX J1947.3+3045 &S&19H47M23.4S & 30D45M55S  &19.2&1.170&0.064&$-$0.96&0.03&$-$0.18&0.68&1RXS J194723.8+304558\\   
RX J1950.0+3821 &A&19H50M04.7S & 38D21M52S  &20.3&0.117&0.023& 0.92&0.08&$-$0.44&0.16&1RXS J195005.2+382155\\   
RX J1951.7+3716 &A&19H51M46.2S & 37D16M56S  &27.5&0.102&0.023& 0.89&0.11&$-$0.42&0.19&1RXS J195148.7+371712\\   
RX J1953.6+5025 &B&19H53M41.0S & 50D25M03S  &22.9&0.203&0.018& 0.11&0.09& 0.01&0.13&1RXS J195340.6+502456\\   
RX J1956.7+5304 &B&19H56M46.7S & 53D04M31S  &31.0&0.170&0.014& 0.45&0.08&$-$0.26&0.10&1RXS J195646.5+530424\\   
RX J2002.1+5438 &B&20H02M07.9S & 54D38M05S  &31.0&0.435&0.023& 0.27&0.05&$-$0.25&0.06&1RXS J200208.2+543827\\   
RX J2010.5+0632 &B&20H10M34.3S & 06D32M06S  &21.5&0.881&0.070&$-$0.20&0.08&$-$0.21&0.11&1RXS J201034.8+063208\\   
RX J2014.8+4501 &B&20H14M49.8S & 45D01M32S  &22.9&0.610&0.030& 0.46&0.05&$-$0.07&0.06&1RXS J201449.4+450143\\   
RX J2019.8+2256 &A&20H19M48.9S & 22D56M21S  &18.4&0.523&0.037&$-$0.21&0.08&$-$0.44&0.11&1RXS J201949.3+225628\\   
RX J2021.7+5213 &A&20H21M44.2S & 52D13M54S  &25.9&0.071&0.011& 0.03&0.37&$-$0.59&0.27&1RXS J202144.2+521348\\   
RX J2033.4+3128 &B&20H33M25.3S & 31D28M03S  &20.3&0.273&0.026& 0.26&0.10& 0.02&0.13&1RXS J203324.9+312816\\   
RX J2040.3+1059 &B&20H40M18.5S & 10D59M40S  &19.2&0.532&0.052& 0.45&0.09&$-$0.05&0.11&1RXS J204019.1+105941\\   
RX J2043.9+5314 &X&20H43M59.4S & 53D14M29S  &21.5&0.229&0.018& 0.95&0.04& 0.47&0.07&1RXS J204400.1+531434\\   
RX J2044.0+2833 &B&20H44M03.9S & 28D33M05S  &19.2&0.303&0.030& 0.91&0.06&$-$0.24&0.09&1RXS J204404.0+283303\\   
RX J2102.0+3359 &S&21H02M02.9S & 33D59M22S  &24.3&0.160&0.019&$-$0.66&0.12&$-$0.76&0.18&1RXS J210201.7+335932\\   
RX J2102.6+4553 &A&21H02M40.5S & 45D53M05S  &24.3&0.216&0.022&$-$0.12&0.10&$-$0.68&0.19&1RXS J210241.4+455305\\   
RX J2104.2+2118 &X&21H04M16.2S & 21D18M20S  &20.3&0.145&0.019& 0.88&0.11& 0.19&0.15&1RXS J210416.4+211816\\   
RX J2109.7+4029 &X&21H09M47.3S & 40D29M47S  &31.0&0.121&0.018& 0.82&0.17& 0.21&0.14&1RXS J210948.0+402944\\   
RX J2133.7+5107 &B&21H33M43.5S & 51D07M20S  &19.2&0.513&0.031& 0.90&0.06&$-$0.16&0.06&1RXS J213344.1+510725\\   
RX J2155.3+5938 &S&21H55M21.0S & 59D38M49S  &34.7&0.112&0.024&$-$0.41&0.18&$-$0.59&0.32&1RXS J215522.8+593843\\   
RX J2156.3+3318 &B&21H56M23.3S & 33D18M40S  &22.9&0.516&0.042& 0.99&0.01&$-$0.23&0.09&1RXS J215623.8+331829\\   
RX J2255.0+5540 &X&22H55M03.3S & 55D40M53S  &24.3&0.103&0.026& 0.75&0.22& 0.04&0.17&1RXS J225504.1+554052\\   
RX J2322.6+6113 &X&23H22M38.3S & 61D13M29S  &34.7&0.118&0.023& 0.87&0.10& 0.06&0.20&1RXS J232241.3+611335\\   
\end{supertabular}

\newpage
\begin{table*}

   \caption[]{Optical identifications of ROSAT survey sources.  The second column
indicates the selection origin of the source as shown on Fig. 6, and given in Table 4.
The last column indicates whether a finding chart is provided.}

\end{table*}
\scriptsize
\tabcolsep=5pt
\tablehead{\hline \tbsp
Source & Sel & d(x-o) & \multicolumn{3}{c}{Right Asce.}     & \multicolumn{3}{c}{Dec}    &
Optical        & V & B & Class & Type & Finding\\[0.8ex]
Name   &     & (r$_{90}$)  & h & m & s &  d & \arcmin & \arcsec   & 
Identification &     &   &    &      & Chart \\[0.8ex]
\hline \tbsp}
\tabletail{\hline \multicolumn{13}{l}{\sl Continued on next page}\\ \hline}
\tablelasttail{\hline}

\begin{supertabular}{crrrrrrrrrrrllc}
RX J0035.8+5950 &X&    *& *& *&    *&  *& *&   *&??                  &    *&    *&UNID &        &Y\\   
RX J0119.0+7033 &X& 0.64& 1&19& 3.32& 70&33&25.7&GSC0430101235       & 9.89&    *&AC   &K1V     & \\   
RX J0143.7+6349 &A& 1.09& 1&43&43.57& 63&49&54.0&HD 10436            & 8.41&    *&AC   &K5V     & \\   
RX J0146.0+6348 &A& 0.31& 1&46& 4.96& 63&48&56.4&HD 10663            & 8.69&    *&AC   &G2V     & \\   
RX J0202.2+6934 &B& 0.09& 2& 2&14.60& 69&34&21.1&GSC0431501020       &11.72&    *&AC   &G2V     & \\   
RX J0222.1+5221 &X& 0.28& 2&22& 6.19& 52&21& 8.4&PMM1350$-$02324516    &    *&17.00&AGN  &Sy 1    & \\   
RX J0222.5+5033 &B& 0.91& 2&22&33.42& 50&33&41.9&BD+49 646           &10.10&    *&AC   &G2V     & \\   
RX J0254.6+3931 &B& 0.59& 2&54&42.63& 39&31&34.7&CJ2 0251+393        &16.10&    *&AGN  &BLRG    & \\   
RX J0256.3+6141 &A& 0.21& 2&56&20.11& 61&41&22.6&GSC0404800621       &13.08&    *&AC   &M2Ve    & \\   
RX J0324.7+3410 &X& 0.66& 3&24&41.16& 34&10&45.9&GSC0234901904       &15.10&    *&AGN  &Sy 1    & \\   
RX J0325.2+4042 &B& 1.82& 3&25&13.22& 40&41&54.6&PMM1275$-$02314407    &15.20&    *&AGN  &Sy 2:   & \\   
RX J0337.0+4738 &X& 0.42& 3&37& 3.14& 47&38&51.5&PMM1350$-$03755404    &    *&17.00&AGN  &Sy 1    & \\   
RX J0416.0+5237 &A& 0.35& 4&16& 3.54& 52&37&37.8&GSC0371900494       &11.70&    *&AC   &K7V     & \\   
RX J0434.7+4014 &X& 0.42& 4&34&41.54& 40&14&19.3&PMM1275$-$03482924    &    *&15.20&AGN  &Sy 1    & \\   
RX J0452.0+4932 &X& 0.09& 4&52& 5.00& 49&32&45.2&H0432052            &17.10&    *&AGN  &Sy 1    & \\   
RX J0459.8+1808 &B& 0.34& 4&59&51.90& 18& 8&38.9&PMM1050$-$01591537    &    *&19.50&AGN  &Sy 1    & \\   
RX J0508.3+1721 &X& 0.71& 5& 8&19.72& 17&21&48.1&GSC0128601162       &13.47&    *&AGN  &Sy 2:   & \\   
RX J0508.9+2113 &X& 0.25& 5& 8&55.15& 21&13& 2.4&PMM1050$-$01718259    &    *&17.70&AGN  &Sy 1    & \\   
RX J0529.0+0934 &X&    *& *& *&    *&  *& *&   *&??                  &    *&    *&UNID &        &Y\\   
RX J0535.0+6450 &S& 0.32& 5&35& 0.62& 64&50&39.1&GSC0408900790       &12.37&    *&AC   &K       & \\   
RX J0554.7+1055 &A& 0.09& 5&54&45.87& 10&55&57.9&GSC0072000052       &11.93&    *&AC   &G0V     & \\   
RX J0602.1+2828 &X& 0.20& 6& 2&10.70& 28&28&22.1&PMM1125$-$03274684    &    *&15.00&AGN  &Sy 1    & \\   
RX J0602.2+2837 &X&    *& *& *&    *&  *& *&   *&??                  &    *&    *&UNID &        &Y\\   
RX J0608.0+3058 &X& 0.64& 6& 8& 0.95& 30&58&42.0&PMM1200$-$04416825    &    *&17.30&AGN  &Sy 1    & \\   
RX J0620.6+2644 &X&    *& *& *&    *&  *& *&   *&??                  &    *&    *&UNID &        &Y\\   
RX J0621.2+4415 &A& 3.05& 6&21&12.50& 44&14&43.4&G 101$-$35            &11.93&    *&AC   &M2Ve    & \\   
RX J0621.7+1747 &X&    *& *& *&    *&  *& *&   *&??                  &    *&    *&UNID &        &Y\\   
RX J0625.8$-$0101 &S& 0.34& 6&25&49.51& $-$1& 1&46.3&GSC0478501175       &12.69&    *&AC   &M1Ve    & \\   
RX J0635.9+0755 &X& 0.35& 6&35&58.28&  7&55&21.0&GSC0073302098       &10.69&    *&AC   &G$-$K     & \\   
RX J0648.7+1516 &X&    *& *& *&    *&  *& *&   *&??                  &    *&    *&UNID &        &Y\\   
RX J0649.8$-$0737 &S& 0.49& 6&49&50.90& $-$7&37&40.2&PMM0750$-$02887019    &    *&18.00&CV   &        &Y\\   
RX J0656.8$-$1424 &X& 0.61& 6&56&54.33&$-$14&24&59.0&GSC0539202173       & 9.77&    *&AC   &K0V     & \\   
RX J0702.0+1257 &S& 0.29& 7& 2& 3.99& 12&57&55.4&GSC0075701608       & 9.89&    *&AC   &Ke      & \\   
RX J0704.5$-$0612 &X& 0.75& 7& 4&31.46& $-$6&12&17.2&GSC0482603053 ??    &13.00&    *&AC?? &G$-$K     & \\   
RX J0713.1$-$0511 &B& 0.53& 7&13&11.22& $-$5&11&43.4&GSC0482302265       &12.14&    *&AC   &M3Ve    & \\   
RX J0717.4$-$1119 &X&    *& *& *&    *&  *& *&   *&??                  &    *&    *&UNID &        &Y\\   
RX J0721.3$-$5720 &B& 0.42& 7&21&23.76&$-$57&20&37.8&GSC0855901016   comp&11.23&    *&AC   &K0V+M5eV& \\   
RX J0749.1$-$0549 &S& 0.47& 7&49&10.45& $-$5&49&24.9&PMM0825$-$05603282    &    *&19.00&CV   &        &Y\\   
RX J0750.9+0320 &B& 1.46& 7&51& 0.73&  3&20&40.9&PMM0900$-$05507668    &    *&15.20&AGN  &Sy 1    & \\   
RX J0755.7$-$0157 &B& 0.36& 7&55&47.32& $-$1&57&41.7&PMM0825$-$05763804    &    *&14.80&AGN  &Sy 1    & \\   
RX J0759.1+0748 &A&    *& *& *&    *&  *& *&   *&??                  &    *&    *&UNID &        &Y\\   
RX J0801.9$-$4946 &X& 0.33& 8& 1&58.00&$-$49&46&36.0&ESO 209$-$ 12         &11.07&    *&AGN  &Sa Sy 1 & \\   
RX J0811.1$-$5555 &B& 0.82& 8&11& 9.37&$-$55&55&56.5&GSC0857001980       &11.97&    *&AC   &G$-$K0    & \\   
RX J0816.4$-$1311 &B& 0.51& 8&16&27.19&$-$13&11&52.6&PMM0750$-$06007988    &    *&17.40&AGN  &BL LAC  & \\   
RX J0818.9$-$2252 &X& 0.28& 8&18&57.70&$-$22&52&35.9&PMM0600$-$05866078    &    *&15.50&AGN  &Sy 1    & \\   
RX J0819.2$-$0756 &A&    *& *& *&    *&  *& *&   *&??                  &    *&    *&UNID &        &Y\\   
RX J0828.5$-$5138 &B& 0.58& 8&28&32.34&$-$51&38&32.6&GSC0816200330       &10.87&    *&AC   &G       & \\   
RX J0845.7$-$3544 &A& 0.48& 8&45&42.51&$-$35&44&36.6&GSC0714900583       &12.09&    *&AC   &G$-$K     & \\   
RX J0856.4$-$2241 &A& 0.10& 8&56&26.41&$-$22&41&39.8&GSC0658500334       &12.41&    *&AC   &M0$-$1V(e)& \\   
RX J0913.0$-$2103 &B& 0.51& 9&13& 0.17&$-$21& 3&20.9&PMM0675$-$07248276    &    *&17.30&AGN  &BL LAC  & \\   
RX J0935.5$-$2802 &B& 0.23& 9&35&31.28&$-$28& 2&54.0&GSC0660900298       &12.08&    *&AC   &M0Ve    & \\   
RX J1023.9$-$4336 &B& 0.35&10&23&56.11&$-$43&36& 2.5&PMM0450$-$11341838    &    *&17.50&AGN  &BL LAC  & \\   
RX J1155.4$-$5641 &B& 0.27&11&55&27.03&$-$56&41&53.3&PMM0300$-$14373754    &    *&13.60&CV   &        &Y\\   
RX J1606.6$-$4618 &X& 0.18&16& 6&36.75&$-$46&18&26.6&GSC0831000256       &12.03&    *&AC   &G$-$K     & \\   
RX J1639.7$-$3920 &X& 0.83&16&39&47.29&$-$39&20&40.3&GSC0785402093       &13.29&    *&AC   &M0Ve    & \\   
RX J1704.3$-$4020 &X& 1.32&17& 4&17.93&$-$40&19&50.2&HD 322763           &10.60&    *&AC   &A3      & \\   
RX J1718.4$-$4029 &X&    *& *& *&    *&  *& *&   *&??                  &    *&    *&UNID &        &Y\\   
RX J1739.4$-$2942 &X& 0.16&17&39&30.30&$-$29&42& 8.9&A                   &    *&    *&Be/X &Be      &Y\\   
RX J1740.7$-$2818 &X&    *& *& *&    *&  *& *&   *&??                  &    *&    *&UNID &        &Y\\   
RX J1741.4+0348 &X& 0.67&17&41&28.26&  3&48&52.9&PMM0900$-$10385820    &    *&15.30&AGN  &Sy 1    & \\   
RX J1742.3$-$2737 &X&    *& *& *&    *&  *& *&   *&??                  &    *&    *&UNID &        &Y\\   
RX J1749.8$-$3312 &X&    *& *& *&    *&  *& *&   *&SLX 1746$-$331        &    *&    *&UNID &        &Y\\   
RX J1804.1+0042 &X&    *&18& 4& 9.01&  0&42&22.1&PMM0900$-$11515260    &    *&18.60&Gal  &        & \\   
RX J1925.0+4429 &A& 0.34&19&25& 1.98& 44&29&50.2&GSC0314600035       & 9.71&    *&AC   &K1V     & \\   
RX J1929.8+4622 &A& 0.38&19&29&50.49& 46&22&23.6&PMM1350$-$10836797    &    *&17.00&AGN  &Sy 1    & \\   
RX J1931.1+0937 &X& 0.39&19&31& 9.23&  9&37&16.4&A                   &    *&19.00&AGN  &Bl Lac  &Y\\   
RX J1935.4+3746 &A& 0.02&19&35&29.40& 37&46& 9.9&GSC0313500052       &11.41&    *&AC   &M4Ve    & \\   
RX J1936.3+2632 &S& 0.53&19&36&18.73& 26&32&56.5&A                   &    *&    *&WD   &DA      &Y\\   
RX J1943.9+2118 &X&    *& *& *&    *&  *& *&   *&??                  &    *&    *&UNID &        &Y\\   
RX J1946.2$-$0444 &A& 0.37&19&46&16.42& $-$4&44&55.4&PMM0825$-$16960702    &    *&19.00&CV   &        &Y\\   
RX J1947.3+3045 &S& 0.66&19&47&24.38& 30&45&55.2&PMM1200$-$13387458    &    *&17.10&WD   &DA      & \\   
RX J1950.0+3821 &A&    *& *& *&    *&  *& *&   *&??                  &    *&    *&UNID &        &Y\\   
RX J1951.7+3716 &A& 0.65&19&51&47.53& 37&16&48.3&PMM1200$-$13720959    &    *&15.50&CV   &        &Y\\   
RX J1953.6+5025 &B& 0.30&19&53&40.76& 50&24&57.2&PMM1350$-$11463233    &    *&15.80&AC   &M3Ve    & \\   
RX J1956.7+5304 &B& 0.30&19&56&46.11& 53& 4&23.2&GSC0393501634       &10.34&    *&AC   &K0V     & \\   
RX J2002.1+5438 &B& 0.61&20& 2& 7.40& 54&37&47.0&HD 190398           & 8.20& 9.40&AC   &G0V     & \\   
RX J2010.5+0632 &B& 0.50&20&10&34.41&  6&32&17.4&GSC0050701588       &12.01&    *&AC   &Me      & \\   
RX J2014.8+4501 &B& 0.72&20&14&49.04& 45& 1&46.4&HD 192785           & 9.20&10.30&AC   &K0V     & \\   
RX J2019.8+2256 &A& 0.79&20&19&49.23& 22&56&35.1&GSC0215501614       &11.40&    *&AC   &M4Ve    & \\   
RX J2021.7+5213 &A& 0.63&20&21&44.59& 52&13&38.6&GSC0358400565       &11.84&    *&AC   &G3V     & \\   
RX J2033.4+3128 &B& 0.73&20&33&24.41& 31&28&12.6&GSC0268600876       &10.35&    *&AC   &G7V     & \\   
RX J2040.3+1059 &B& 0.24&20&40&18.55& 10&59&45.2&PMM0975$-$19711620    &    *&16.10&AGN  &Sy 1    & \\   
RX J2043.9+5314 &X& 0.26&20&43&59.60& 53&14&34.9&PMM1425$-$11257915    &    *&17.50&AGN  &Sy 1    & \\   
RX J2044.0+2833 &B& 0.52&20&44& 4.50& 28&33&12.1&PMM1125$-$17121513    &    *&14.70&AGN  &Sy 1    & \\   
RX J2102.0+3359 &S&    *& *& *&    *&  *& *&   *&?? UV excess object &    *&    *&UNID &        &Y\\   
RX J2102.6+4553 &A& 0.84&21& 2&38.67& 45&52&58.4&HD 200560           & 7.68& 8.65&AC   &K3V     & \\   
RX J2104.2+2118 &X&    *& *& *&    *&  *& *&   *&??                  &    *&    *&UNID &        &Y\\   
RX J2109.7+4029 &X& 0.87&21& 9&48.50& 40&29&23.9&GSC0317201505       &10.70&    *&AC   &G2V     & \\   
RX J2133.7+5107 &B& 0.21&21&33&43.64& 51& 7&24.4&PMM1350$-$14702073    &    *&15.80&CV   &        &Y\\   
RX J2155.3+5938 &S& 0.69&21&55&23.70& 59&38&36.6&PMM1425$-$12649991    &    *&12.70&AC   &M4Ve    & \\   
RX J2156.3+3318 &B& 0.24&21&56&23.04& 33&18&35.9&PMM1200$-$18997159    &    *&16.20&Gal  &        & \\   
RX J2255.0+5540 &X& 0.44&22&55& 4.57& 55&40&52.5&GSC0398901784       & 9.42&    *&AC   &G0V     & \\   
RX J2322.6+6113 &X& 0.38&23&22&40.08& 61&13&33.5&GSC0427901821       & 9.79&    *&AC   &G0V     & \\   
\end{supertabular}

\end{document}